\documentclass[twocolumn,floatfix,showpacs,amsmath,amssymb,aps,prb]{revtex4}

\usepackage{graphicx}
\usepackage{dcolumn}
\usepackage{bm}
\usepackage{ulem,color,verbatim}
\renewcommand{\emph}[1]{\textit{#1}} 
\usepackage{color}

\begin{document}

\title{Kinks in the electronic dispersion of the 
Hubbard model away from half filling}

\author{Patrick Grete, Sebastian Schmitt, Carsten Raas, Frithjof B. Anders, 
and G\"otz S. Uhrig}
\affiliation{Technische Universit\"at Dortmund, 44221 Dortmund, Germany}
\email[E-Mail:]{sebastian.schmitt@tu-dortmund.de}

\date{\today}

\begin{abstract}
We study kinks in the electronic dispersion of a generic strongly correlated 
system by dynamic mean-field theory (DMFT). The focus is on doped systems away 
from particle-hole
symmetry where valence fluctuations matter potentially. Three different 
algorithms are compared to asses their strengths and weaknesses as well as to 
clearly distinguish physical features from algorithmic artefacts.
Our findings extend a view previously established for 
half-filled systems where kinks reflect the coupling
of the fermionic quasiparticles to emergent collective modes, 
which are identified here as spin fluctuations.
Kinks are observed when strong spin fluctuations are present and, 
additionally, a separation of energy scales for spin and charge excitations 
exists. Both criteria are met by strongly correlated systems close to a 
Mott-insulator transition.
The energies of the kinks and their doping dependence 
fit well to the kinks in the cuprates, which is surprising
in view of the spatial correlations neglected by DMFT.
\end{abstract}




\maketitle

\section{Introduction}

Collective bosonic modes can modify the   low-energy electronic
properties on the characteristic energy scale of these bosonic excitations.
Well understood are the significant mass enhancement\cite{AshcroftWilkins1965} 
and the kinks in the electronic dispersion\cite{Mahan81} in materials with 
strong electron-phonon coupling. These kinks are abrupt changes of the slope 
of the electronic dispersion which occur
at energies of the order of the Debye energy which is much smaller than the 
Fermi energy in such systems.

In this paper we study kinks in the electronic dispersion
which are related to emergent collective bosonic modes 
of the purely electronic system  rather than by a coupling to external 
bosons such as phonons. 
It was shown by Byczuk et al. in the framework of dynamic mean-field
theory\cite{pruschke:dmftNCA_HM95,georges:dmft96} (DMFT) that such
kinks are a generic feature of strongly interacting systems and require no 
externally coupled bosons.\cite{byczukKinks09} Their argument resides
on the structure of the DMFT equations and the spectral density of the 
local propagator.

Our work supplements
the mathematical argument in Ref.\ \onlinecite{byczukKinks09} by the physical 
picture that the \textit{internally generated} emergent collective modes 
provide the bosons which generate the kinks in the dispersion.
This point of view has been already put forward for the half-filled case  
where the physical situation is particularly clear.\cite{Raas2009}
We will show that in the model studied below these 
collective modes are spin fluctuations of local moments formed 
due to the strong local Coulomb interaction.
In  the Mott insulating phase, stable local moments are generated which tend 
to form magnetically ordered phases due to residual spin-spin interactions.
In the metallic phase, however, these local moments are completely screened at 
zero energy and the spin fluctuations at intermediate energies live only for
a finite time because they decay into particle-hole pairs (Landau damping).
We call the spin fluctuations appearing as resonances emergent
modes because they become long-lived for even larger interaction in the 
insulating phase. Photoemission experiments support our view in 
antiferromagnetic\cite{kordy04,kordy06,dahm09} 
and in  ferromagnetic systems.\cite{hofma09}

As will be shown, there exist two conditions for the kinks 
to appear in the dispersion relation. For one, strong emergent 
collective modes (spin fluctuations) must be present in the system. 
This is signaled by   a pronounced maximum in the imaginary part of 
the spin susceptibility at a finite energy $\omega_{\text{max}}$. 
For another, the energy scale for charge excitations $\omega_{\text{charge}}$
must be much larger than the characteristic spin-excitation energy, i.e.\ 
$|\omega_{\text{charge}}|\gg|\omega_{\text{max}}|$.
Here, the energy scale for charge excitations is set
either by  a single-particle energy or  the characteristic scale for collective 
charge excitations, depending on which is smaller.
Both criteria are met in strongly correlated metallic systems close to a 
Mott-insulator transition.

The term ``collective bosonic modes'' in solid-state systems usually refers to 
dispersive, i.e., wave-vector dependent, excitations such as (para-)magnons 
or charge density waves.  Such modes are characterized by the 
momentum dependent spin- and charge-susceptibilities 
$\chi_{\textrm{mag}}(\vec q,\omega)$ and  
$\chi_{\textrm{charge}}(\vec q,\omega)$, respectively.
It may be put forward that in DMFT no true collective modes exist
because they do not propagate properly. Indeed, in the strict limit of infinite
dimensions collective two-particle properties are local because
they are momentum-independent except for some momenta of measure zero.
\cite{mulle89a,donge89,uhrig93b,uhrig95d} Yet we do not adopt
this puristic view and stick to the wording of
collective bosonic modes for three reasons. First, spin- and charge
fluctuations are two-particle properties and as such comprise
at least two elementary fermionic excitations so that they are
collective and bosonic in this sense. Second, even in the strict
DMFT, the spin- and charge-susceptibilities are true lattice quantities
involving all lattice sites. Otherwise, their imaginary parts
would show only one or two $\delta$ peaks instead of broad
continua, see for instance Ref.\ \onlinecite{Raas2009a}.
Third, generically the DMFT is taken as an approximation
to finite dimensional systems. Then, the susceptibilities acquire a 
non-trivial  $\vec q$-dependence through the inner momentum and frequency 
sums in the Bethe-Salpeter equations, 
even though the irreducible two-particle vertex is assumed to be local. 
\cite{jarrellQMCHM92,schmittPhD08}.
In the present work, we extract the characteristic energy scales relevant for 
the dispersive collective modes from the local susceptibilities 
$\chi(\omega)=\frac 1N\sum_{\vec q}\chi(\vec q,\omega)$
because they readily reflect these energy scales and can
be obtained in DMFT from the effective impurity model.

We investigate the low-temperature phase of the Hubbard model away from half 
filling using the  DMFT with three different impurity solvers in order to
be able to clearly separate physical features from possible algorithmic 
artifacts. We employ two numerical renormalization schemes, namely the 
density-matrix renormalization group\cite{schollwoeckDMRG05}
(DMRG) and numerical renormalization group\cite{bullaNRGReview08} (NRG)
as well as    a diagrammatic approach, the so-called
enhanced non-crossing approximation (ENCA).\cite{pruschkeENCA89,keiterENCA90}
The results of all three algorithms agree
very well. The remaining differences can be understood on the basis of the
known strengths and weaknesses of the approaches.

The prevailing effect  of the interaction is a significant
renormalization of the bare electron mass due to correlations.
The Fermi liquid theory\cite{nozie97} connects this mass renormalization
to a fundamental energy scale $T^*$ below which the Fermi liquid picture of 
renormalized weakly interacting,  long-lived quasiparticles
applies. In strongly correlated electron systems  
the properties of these renormalized  quasiparticles may differ
dramatically from those of non-interacting electrons as exemplified in
heavy fermion systems.\cite{greweSteglichHF91,HeavyFermionsReviewColeman07}

The article is laid out as follows.
The model and the relevant theoretical background  are described 
Sec.\ \ref{sec:model}.
In particular, the basic ideas of all three impurity solvers are discussed.
In order to gauge their differences a comparison of the  
single-particle dynamics  is shown in Sec.\ \ref{sec:methods}. 
We  begin with the results for the self-energies. Then we turn 
to the extraction of  the Fermi liquid low energy scale $T^*$
and relate it to the maxima in the local charge and spin susceptibilities.
in Sec.\ \ref{sec:low-energy-scale}. We will explicitly show that $T^*$ 
also occurs as energy scale in the dynamic  spin susceptibility
but not in the local charge  susceptibility.  Therefore, 
the connection made between emergent spin-fluctuations and the
the kinks in the electronic dispersion\cite{Raas2009} extends to
the metallic regime away from half filling.
This main results of our work is contained in
Sec.\ \ref{sec:kinks-collective-modes} 
where also  the  doping dependence of the kink energies is compared to  
experimental results.
Finally, our findings are summarized in Sect.\ \ref{sec:summary}.

\section{Model and methods}
\label{sec:model}
\subsection{Hubbard Model and Dynamic Mean-Field Theory}

We consider the single-band Hubbard model which is the simplest model for 
correlated electrons on a lattice
\begin{align}
  \label{hubham}
  \begin{split}
    \mathcal{H} = 
    &- t \sum_{\left<i,j\right>,\sigma}
    (c_{i\sigma}^\dagger c_{j\sigma}^{\phantom{\dagger}}
    +c_{j\sigma}^\dagger c_{i\sigma}^{\phantom{\dagger}})\\
    &+ U \sum_i
    n_{i\uparrow}n_{i\downarrow}
    - \left(\frac{U}{2} + \mu\right)
    \sum_{i\sigma} n_{i\sigma}
    .
  \end{split}
\end{align}
The operators $c^{\dagger}_{i\sigma}$ ($c^{\phantom{\dagger}}_{i\sigma}$) 
create (annihilate)  electrons with spin $\sigma =\{\uparrow,\downarrow\}$ 
at lattice sites $i$ and $n_{i\sigma}=c^{\dagger}_{i\sigma}c^{\phantom{\dagger}}_{i\sigma}$ 
is the occupation number operator. 
The first term in the Hamiltonian~\eqref{hubham} 
describes electronic hopping with amplitude $t$,
where $\left<i,j\right>$ indicates nearest neighbor sites. 
The local Coulomb repulsion is incorporated in the second term with 
matrix element $U$.
The last term sets the local single-particle levels and includes the chemical 
potential in a way, that $\mu=0$ corresponds to an electron-hole symmetric 
situation, i.e., half-filling. 

Despite its simplicity, the exact solution of the Hubbard 
Hamiltonian~\eqref{hubham} has only been possible in one spatial
dimension,\cite{liebWu:HubbardModel,lieb1DHubbard03} for a recent book on 
this topic see Ref.~\onlinecite{Essler2005}. 
In order to obtain an approximate solution for  higher  dimensional systems
we employ the dynamic mean-field theory 
(DMFT),\cite{pruschke:dmftNCA_HM95,georges:dmft96}
for recent reviews see Refs.~\onlinecite{georgesDMFTreview04,kunesDMFT10}.

A non-trivial, but considerable, simplification is obtained in the 
limit of infinite coordination number (infinite spatial dimension) 
if the hopping matrix elements are rescaled appropriately. 
\cite{metznerDInfty89,muellerHartmannDInfty89,georges:HMInfDim92}
In this limit DMFT represents the exact solution.
\cite{pruschke:dmftNCA_HM95,georges:dmft96}
When applied in finite dimensions, the major approximation of DMFT 
consists in treating all non-local correlations in a mean-field manner
while the correlated local dynamics is faithfully retained.
This translates to the self-energy being 
momentum independent,\cite{muellerHartmannDInfty89}
$\Sigma(\vec{k},z)\overset{\mathrm{DMFT}}{\to}\Sigma(z)$ where we use $z$ as a
general complex energy argument with finite imaginary part.

Then, the local Green function of the lattice problem reads
\begin{equation}
  \label{eq:lattsum}
  G(z)=\frac{1}{N}\sum_{\vec{k}}\frac{1}{ z+\mu+\frac U2 -\epsilon_{\vec{k}}
    -\Sigma(z)}
\end{equation}
where $N$ is the number of lattice sites and $\epsilon_{\vec{k}}$ the bare 
dispersion. This local lattice propagator equals the local Green function 
of an effective single-impurity Anderson model (SIAM)
\begin{equation}
  G(z) = \frac{1}{ z+\mu+\frac U2 -\Gamma(z)-\Sigma(z)}
  \label{eq:localGf}
\end{equation}
embedded in a medium characterized by the hybridization function 
$\Gamma(z)$. Thus, the embedding medium is a dynamic medium and
it is not independent from the solution of the
SIAM, but it has to be determined self-consistently as in any generic
mean-field approach.

For a given guess for the hybridization function
the local Green function is determined by a suitable numerical algorithm
which is commonly referred to as the employed `impurity solver'.
This yields the self-energy
\begin{equation}
  \label{eq:dmftSig}
  \Sigma(z)=z+\mu+\frac U 2-\Gamma(z)-G(z)^{-1} \quad,
\end{equation}
which in turn is used to obtain the local lattice propagator $G(z)$ 
via Eq.~\eqref{eq:lattsum}.
It is in this step that the lattice structure enters.  
The self-consistency cycle is
closed by reorganizing Eq.~\eqref{eq:dmftSig} and to extract a new guess for 
$\Gamma(z)$. This cycle is iterated until convergence is reached in the 
pragmatic sense of a tolerable deviation of two successive results for the
local Green function, the self-energy, or the hybridization function.

The nontrivial step in this cycle is the solution of the effective SIAM.
Due to the long history of DMFT, a multitude of different impurity solvers for 
treating this model exists such as iterative perturbation
theory,\cite{georges:HMInfDim92,Kajueter1996,Radzimirski2007} exact
diagonalization\cite{Caffarel1994,Si1994} or several variations of quantum
Monte-Carlo 
schemes.\cite{jarrellQMCHM92,Feldbacher2001,Rubtsov2005,wernerContinousTimeQMC06,GulletAl2011}
 In this work, we employ the enhanced non-crossing approximation, the numerical
renormalization group, and the dynamic density-matrix renormalization
group as impurity solvers and compare their results.

\subsection{Enhanced non-crossing approximation}
\label{sec:enca}

The enhanced non-crossing approximation
(ENCA),\cite{pruschkeENCA89,greweCA108,holmFiniteUNCA89,keiterENCA90} sometimes
also called one-crossing approximation,\cite{hauleDMFTLDA10} is a
thermodynamically conserving 
\cite{baymKadanoffConservation61,baym:conservingApprx62}
approximation for the SIAM which utilizes the expansion with respect to the
hybridization between the impurity electrons and the conduction
band.\cite{keiter:PerturbJAP71,keiter:PerturbIJM71,grewe:IVperturb81,keiterMorandiResolventPerturb84} It extends the usual non-crossing approximation
(NCA)\cite{Grewe:siam83,Kuramoto:ncaI83} to finite values of the Coulomb
repulsion $U$ via the incorporation of the lowest order vertex corrections,
which are necessary to produce the correct Schrieffer-Wolff exchange coupling
and the order of magnitude of the low energy Kondo scale $T_\mathrm{K}$ in the 
problem. The impurity spectral function\cite{pruschkeENCA89} and dynamic
susceptibilities\cite{schmittSus09} are extracted directly for real frequencies
without any adjustable parameters.

The NCA is known to violate Fermi liquid properties at very low temperatures
and some pathological structure appears at the Fermi level below a pathology
scale.\cite{kuramoto:AnalyticsNCA85,bickers:nca87a}
The ENCA removes the cusps associated with this pathology\cite{pruschkeENCA89}
and significantly improves the Fermi liquid properties of the spectral
functions and dynamic susceptibilities.\cite{schmittSus09,schmittPhD08} 
However, the
skeleton diagrams selected within the ENCA still suffer from an imbalance
between charge and spin excitations. While magnetic properties, i.e., the
magnetic susceptibility, are described excellently, charge fluctuations are not
as well accounted for.\cite{schmittSus09} In the spectral functions, where spin
and charge fluctuations contribute equally, this leads to an overestimation of 
the height of the many-body resonance at the Fermi level at too low
temperatures. The dynamic charge susceptibility is overestimated for very low
frequencies. These shortcomings are related to the threshold exponents of the 
auxiliary ionic propagators and 
tend to be worse in parameter regimes with substantial
valence fluctuations, i.e., in the mixed valence regime or at small Coulomb
repulsions $U$.

Within the DMFT the overestimation of the many-body resonance of the impurity
solution might lead to a violation of causality in the self-energy due to the
subtraction occurring in Eq.~(\ref{eq:dmftSig}).  In parameter regimes with 
considerable valence fluctuations, i.e., at small $U$ or at large doping, this 
fact limits DMFT calculations to temperatures above the characteristic low 
temperature scale $T^*$ of the lattice. Because $T^*\to 0$ on approaching the 
Mott-Hubbard metal-to-insulator
transition (MIT) the ENCA can be used as impurity solver down to
very low temperatures in the vicinity of the MIT.

Detailed comparisons of various approximations based on the hybridization
expansion can be found in the literature, e.g., in
Refs.~\onlinecite{pruschke:dmftNCA_HM95},  \onlinecite{greweCA108}, 
\onlinecite{jarrell:HM93b}, and \onlinecite{costi:NCAvsNRG96}.

\subsection{Numerical renormalization group}
\label{sec:nrg}

The NRG is a very powerful tool for accurately calculating equilibrium
properties of complex quantum impurities. Originally developed for
treating the single-channel Kondo Hamiltonian,\cite{wilsonNRG75} this
non-perturbative approach was successfully extended to the Anderson impurity
model\cite{krishnamurtyNRGSIAMI80,krishnamurtyNRGSIAMII80} and other more
complex  quantum impurities.\cite{bullaNRGReview08} At the heart of this
approach is a logarithmic discretization of the continuous conduction band, 
controlled by the discretization parameter\cite{wilsonNRG75} $\Lambda > 1$. 
Using an
appropriate unitary transformation, the discretized Hamiltonian is mapped onto a
semi-infinite chain, with the impurity coupled to the open end. The $N$th link
along the chain represents an exponentially decreasing energy scale, $\omega_N
\propto \Lambda^{-N/2}$. Using this hierarchy of scales, the sequence of
finite-size Hamiltonians $\mathcal{H}_N$ for the $N$-site chain is solved
iteratively, discarding the high-energy states at the end of each step
to maintain a manageable number of states. The reduced basis set of
$\mathcal{H}_N$ thus obtained is expected to faithfully describe the spectrum of
the full Hamiltonian on a scale of $\omega_N$, corresponding to the temperature
$T_N \sim \omega_N$.\cite{wilsonNRG75} 
Because the thermal Boltzmann factors suppress the contributions of higher lying
energy states exponentially, the reduced NRG basis set of $\mathcal{H}_N$ is 
sufficient
for an accurate calculation of thermodynamic quantities at temperature $T_N$.

Dynamical quantities, however, such as impurity Green functions and 
susceptibilities, require the information on all energy scales. In a recent 
extension of the NRG to real-time dynamics out of
equilibrium\cite{andersRealTimeDynNRG05,andersRealTimeKondo06} a complete basis
set for a Wilson chain of length $N$ has been identified which is also used for
the accurate calculation of spectral
functions.\cite{petersNewNRG06,WeichselbaumDelft2007} Additionally, the
discretization error is reduced by averaging over several different
discretizations of the conduction band.\cite{YoshidaWithakerOliveira1990} In
order to reduce the arbitrariness in the spectral
broadening\cite{bulla:NRGSIAM98,bulla:MITFTNRG01,AndersTCSIAM2005} the
single-particle self-energy entering the DMFT equations is obtained from an
exact expression of a ratio of two correlations
functions.\cite{bulla:NRGSIAM98} Since the local dynamic bosonic spin and
charge susceptibilities are calculated directly from the NRG raw
data\cite{petersNewNRG06} more pronounced broadening artifacts occur.

In this work we use a discretization parameter $\Lambda=2$ and keep
approximately $1700$ states in each NRG iteration step. Eight different band
discretizations\cite{YoshidaWithakerOliveira1990} are averaged and the
artificial logarithmic
broadening\cite{bulla:NRGSIAM98}  varies between $b=0.08$ and $b=0.2$.

\subsection{Dynamic density matrix renormalization}

The  DMRG  introduced by White\cite{whiteDMRG92,whiteDMRG93} 
in 1992 is an excellent numerical method for one-dimensional 
quantum systems with open boundary conditions 
\cite{schollwoeckDMRG05,schollwoeckDMRG10} 
such as the SIAM in linear chain representation.
The dimension of the Hamilton matrix grows exponentially with system size. The
DMRG provides a well controlled procedure to cut off this growth by
selecting an optimum basis set for the desired states, 
e.g., the ground state or another target state.
The optimum basis states are selected from the eigenvectors of a reduced
density matrix from which only the largest eigenvalues are retained.

We  calculate dynamic quantities at zero temperature such as the 
advanced Green functions $G(\omega-\mathrm{i}\eta)=G^>(\omega-\mathrm{i}\eta)
+G^<(\omega-\mathrm{i}\eta)$ and $G(\omega)=
\lim_{\eta\to 0+}G(\omega-\mathrm{i}\eta)$  using
\begin{equation}
  \label{greenfuncenergy}
  G^{\gtrless}(\omega-\mathrm{i}\eta):=
  \left<0\left|
      \mathcal{A}\frac{1}{\omega-\mathrm{i}\eta\mp\Delta\mathcal{H}}
      \mathcal{A}^\dagger \right|0\right>,
\end{equation}
where $\Delta\mathcal{H}:= \mathcal{H}-E_0$. The imaginary part of 
$G^{\gtrless}(\omega-\mathrm{i}\eta)$ provides the spectral densities which we 
are aiming at. Several variants of numerical  methods were introduced to obtain
dynamic quantities, for instance the Lanczos method\cite{Garcia2004} and the 
correction vector method.\cite{Ramasesha1997,Kuehner1999,Jeckelmann2002} Since
the Lanczos method has a limited numerical 
resolution,\cite{Garcia2004,Garcia2007a,Garcia2007b,Miranda2008} 
for details see the analysis in Ref.~\onlinecite{raas05}, we use the
correction vector method which targets the ground state
$\left|0\right>$, the excited state $\left|\mathcal{A}\right>:= 
\mathcal{A}^\dagger\left|0\right>$, and the resolvent applied to the excited 
state. This additional targeted state $\left|\xi_\pm\right>$ is called the 
correction vector
\begin{equation}
  \left.
    \left|\xi_\pm\right>\right. :=
  \frac{1}{\omega-\mathrm{i}\eta\pm\Delta\mathcal{H}}\left.
  \left|\mathcal{A}\right>
  \right. .
\end{equation}
Technically, one targets both the real and the imaginary part of 
$\left|\xi_\pm\right>$.
The Green function $G^<(\omega-\mathrm{i}\eta)$ is obtained from the scalar 
product
\begin{equation}
  G^<(\omega-\mathrm{i}\eta)=\left<\mathcal{A}\left|\xi_\pm\right>\right.
\end{equation}
for discrete complex frequencies $\omega_j-\mathrm{i}\eta_j$, $\eta_j>0$. 
In order to
obtain $G(\omega)$ at the real axis with 
continuous spectral density $\rho(\omega):=\mathrm{Im}G(\omega)/\pi$, 
we use the least-bias deconvolution algorithm.\cite{Raas2005}

The key advantage of the correction vector DMRG
is a good energy resolution for low \textit{and} high frequencies
$\omega$. With correction vector DMRG combined with least-bias deconvolution,
local Green functions\cite{Karski2005,Karski2008} and  local
susceptibilities\cite{Raas2009a} have been computed successfully.

In this work, 
we use a fixed distance $\Delta\omega=0.1D$ between two successive frequencies 
$\omega_j$ and $\omega_{j+1}$, the artificial broadening is $\eta_j=0.1D$. 
The energy scale $D$ is half the bandwidth. We keep 256 states in the reduced 
density matrix.

\section{Comparison of Methods}
\label{sec:methods}

In this section, we present results for the Hubbard model within DMFT using the
three impurity solvers introduced in the previous section. 
The non-interacting density of states (DOS) is given by the semi-ellipse
$\rho_0(\omega)=(2/(D\pi))\sqrt{1-(\omega/D)^2}$.

The  NRG and DMRG  calculations were done at zero
temperature while the ENCA requires a small finite $T$ as discussed in
Sect.\ \ref{sec:enca}; the used values are given for each result below.
The spectral density $\rho(\omega)$ of the Green function is
given by $\rho(\omega)=\mathrm{Im}G(\omega)/\pi$
and we similarly define 
$\Sigma(\omega) := \lim_{\eta\to0+}\Sigma(\omega-\mathrm{i}\eta)$.

\subsection{Single Particle Dynamics}
\label{sec:comparison}


\begin{figure}[ht]
  \centering
  \includegraphics[width=\columnwidth]{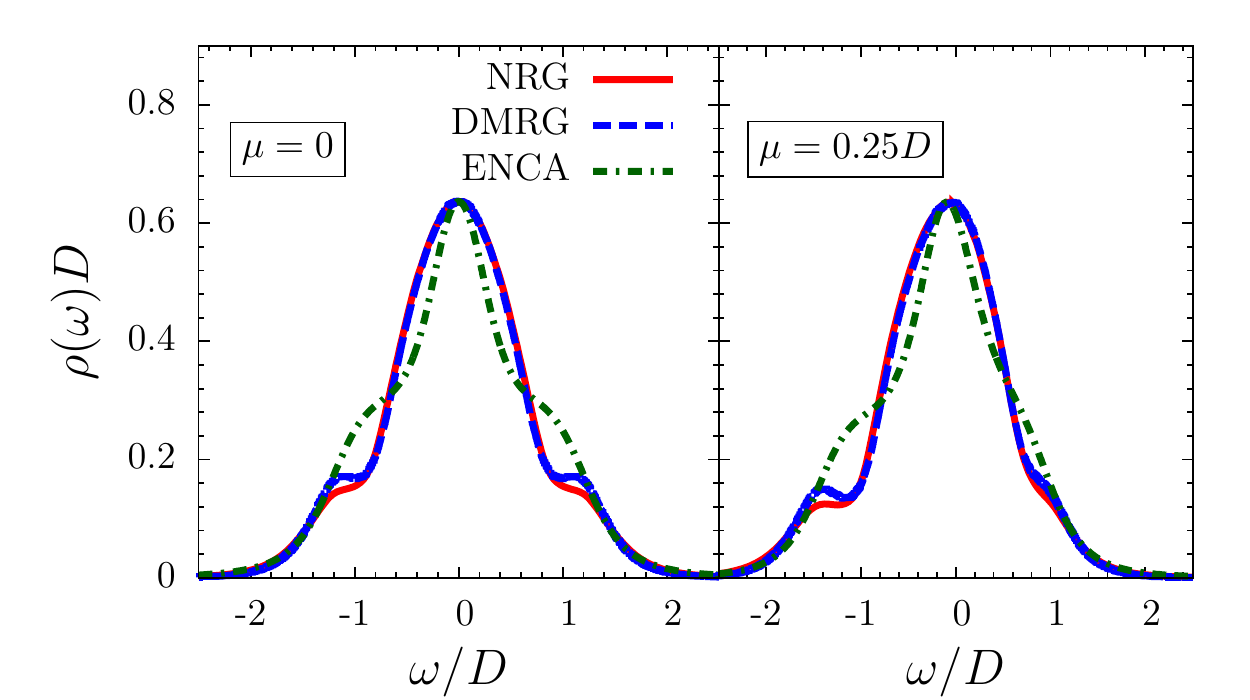}
  \caption{(Color online)
    Spectral function  for $U=D$ and $\mu=0$ (left panel) and $\mu=0.25D$ 
    (right panel).  While the NRG and DMRG results are for $T=0$, 
    the ENCA curves are computed for $T=0.17D$ 
    (left panel) and for $T=0.22D$ (right panel). 
    The case $\mu=0$ implies half-filling, $n=0.5$, while  for $\mu=0.25D$
    the fillings are $n_{\mathrm{NRG}}\approx 0.583$, 
    $n_{\mathrm{DMRG}}\approx 0.577$, and $n_{\mathrm{ENCA}}=0.573$.
  }
  \label{fig:gf_u1}
\end{figure}

\begin{figure}[ht]
  \centering
  \includegraphics[width=\columnwidth]{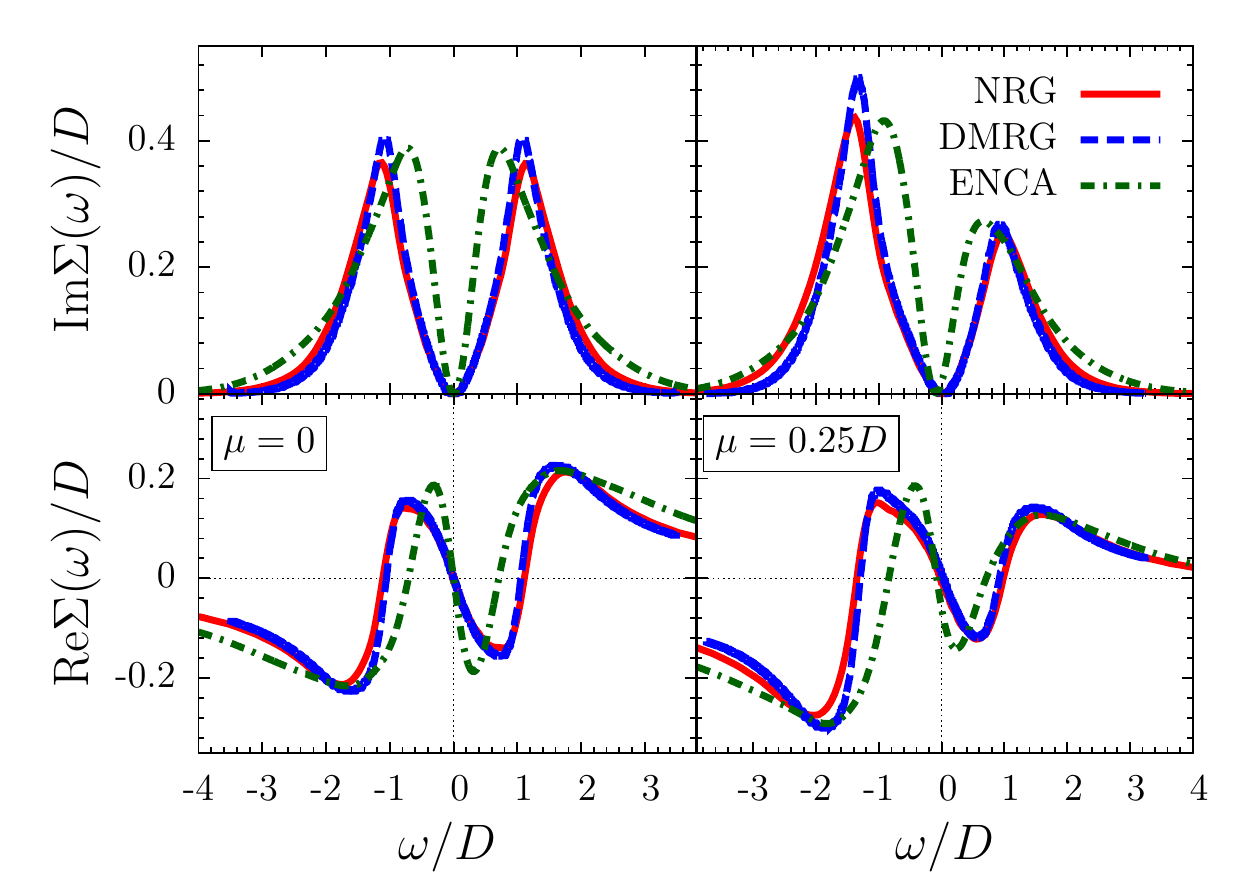}
  \caption{(Color online)
    Imaginary (top row) and real (bottom row) parts of the self-energy for 
    $\mu=0$ (left column) and $\mu=0.25D$ (right column) and $U=D$. 
    Other parameters are as in Fig.~\ref{fig:gf_u1}
  }
  \label{fig:sig_u1}
\end{figure}

Figure~\ref{fig:gf_u1} displays a comparison of the local spectral densities 
for a moderate interaction $U=D$ and for two different values of the chemical 
potential. All three methods qualitatively yield  the same result with a broad 
central peak and only very small shoulders at energies $|\omega|\approx D$.
The latter are precursors of the Hubbard bands centered at the
energies $\omega\approx \pm U/2-\mu$. The
NRG and DMRG results agree quantitatively and only the shoulders are slightly 
more  pronounced in the DMRG curve which is probably due to the lower resolution
of the NRG at high energies.  
The central resonance of the ENCA curve is narrower and the shoulders of the
Hubbard bands are more washed out. The latter feature can be attributed to the
rather high temperature, 
$T\approx 0.17D$ for $\mu=0$ and $T\approx0.23D$ for $\mu=0.25D$,
required to avoid the ENCA problems at too low temperatures, see
Sect.\ II.B. At large energies, e.g., $|\omega+\mu|\gtrsim 1.5D$, the spectra 
of all three methods agree almost perfectly.

The corresponding self-energies are shown in Fig.~\ref{fig:sig_u1}. 
For all methods the imaginary part $\mathrm{Im}\Sigma(\omega)$ displays a 
quadratic minimum at the Fermi level signalling the validity of a
low-energy effective  Fermi liquid description.
This implies that the central peak in the spectral 
function of Fig.~\ref{fig:gf_u1} is essentially due to 
Fermi liquid quasiparticle excitations. In accord with the Kramers-Kronig (KK)
relation, $\mathrm{Re}\Sigma(\omega)$ behaves linearly at the Fermi level.
The self-energies of  the renormalization group (RG) approaches NRG and DMRG
agree quantitatively and the visible deviations are only  due to 
the different broadening procedures used  to obtain continuous functions
in $\omega$.
The ENCA approach yields a too steep quadratic and linear 
dependence around the Fermi level in the imaginary and real part, respectively.

\begin{figure}[htb]
  \centering
  \includegraphics[width=\columnwidth]{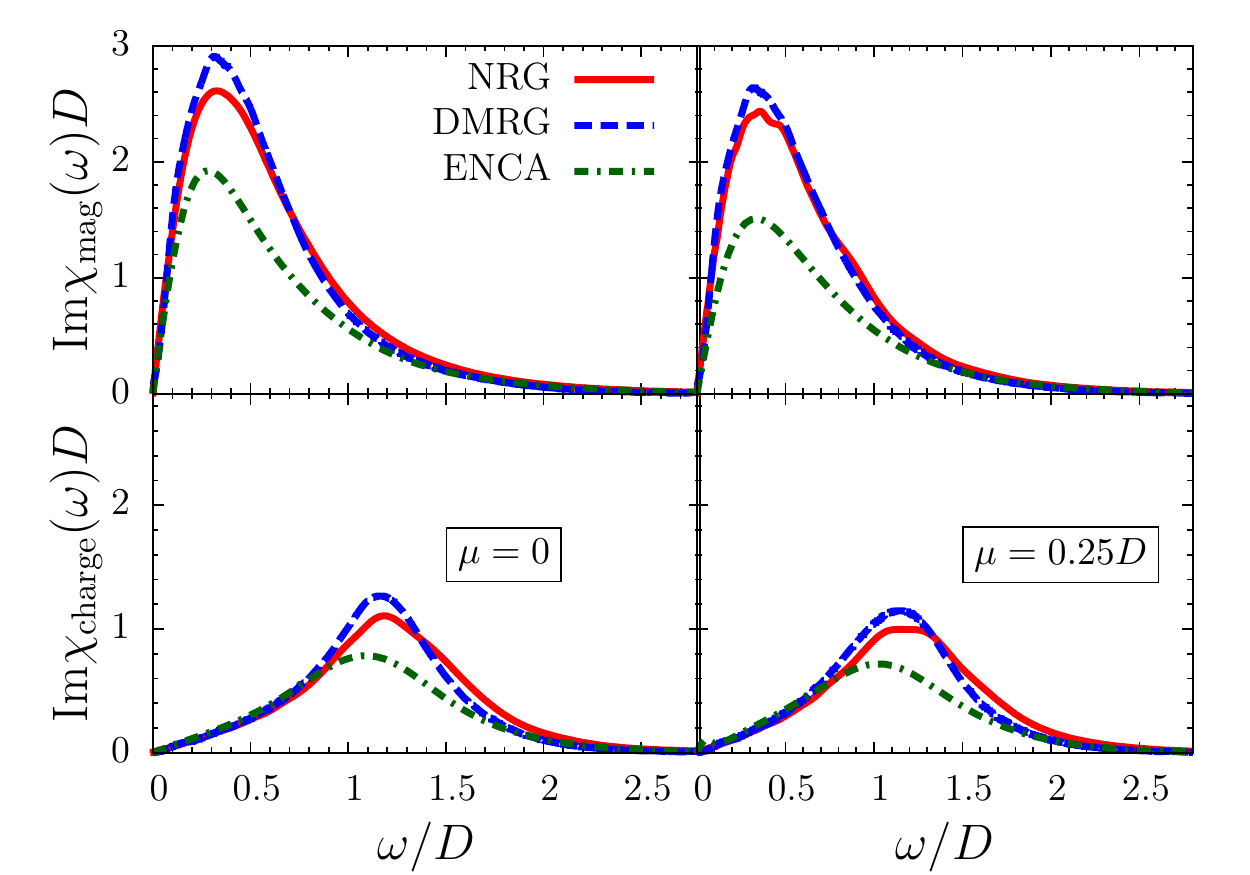}
  \caption{(Color online)
    Imaginary part of the magnetic (top row) and charge (bottom row) 
    susceptibility for $\mu=0$ (left column) and $\mu=0.25D$ (right column) 
    and $U=D$. Other parameters are as in Fig.~\ref{fig:gf_u1}.}
  \label{fig:sus_u1}
\end{figure}

\subsection{Collective Modes and Low Energy Scale}
\label{sec:low-energy-scale}
The imaginary part of the local dynamic magnetic and charge susceptibilities 
shown in Fig.~\ref{fig:sus_u1} shed light on the characteristic
energies for both types of collective excitations.
The characteristic energy for local charge excitations is given by the position
of the Hubbard bands as can be observed in Fig.~\ref{fig:gf_u1}. 
Consequently $\mathrm{Im}\chi_\mathrm{charge}(\omega)$ has a broad peak at 
roughly $\omega\approx 1.1D$.  
Away from half-filling, the asymmetric position  of the lower and the upper 
Hubbard band is reflected by a slightly broadened peak.
Compared to the charge excitations, the characteristic energy for 
local spin excitations is shifted towards lower values and the absolute height 
of $\mathrm{Im}\chi_\mathrm{mag}(\omega)$ is roughly twice as large as 
$\mathrm{Im}\chi_\mathrm{charge}(\omega)$. 
Both these features are signs of the  enhanced magnetic Kondo-like 
correlations in the system 
already present for this moderate value of $U=D$. The overall height of the 
ENCA susceptibilities is lower than those of  the RG approaches due to the 
nonzero temperature.  Additionally, the maxima in the ENCA
results are shifted to slightly lower energies.

\begin{figure}[htb]
  \centering
  \includegraphics[width=\columnwidth]{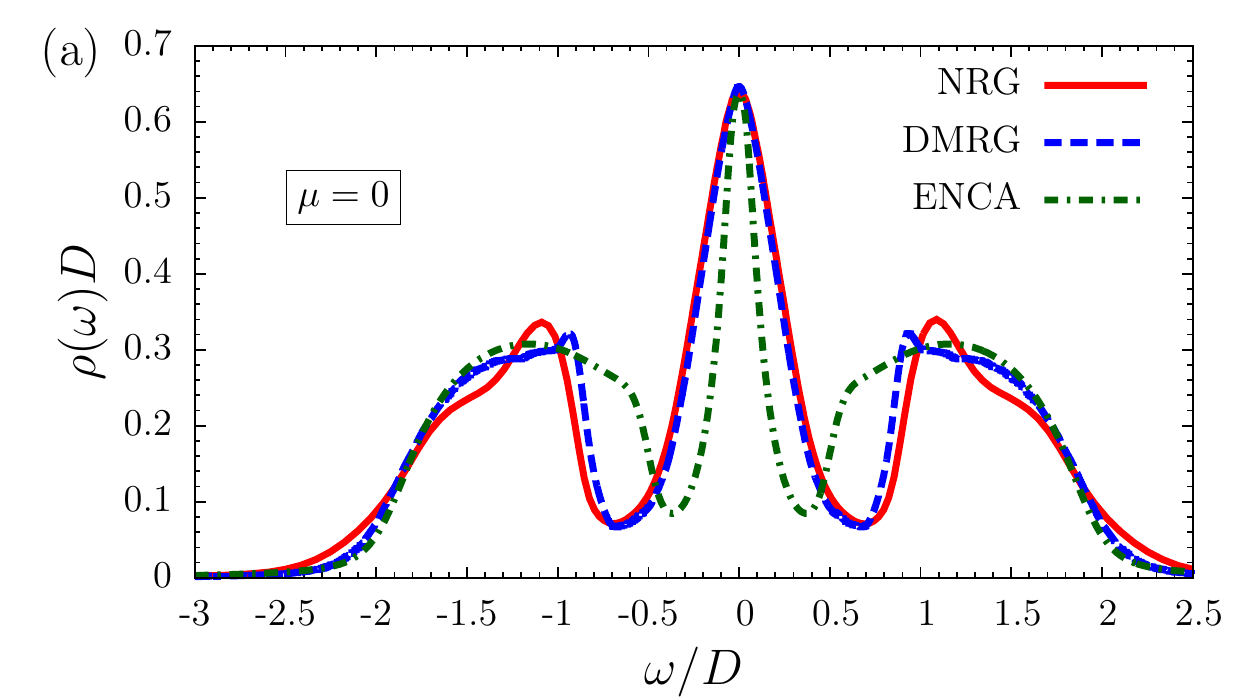}\\
  \includegraphics[width=\columnwidth]{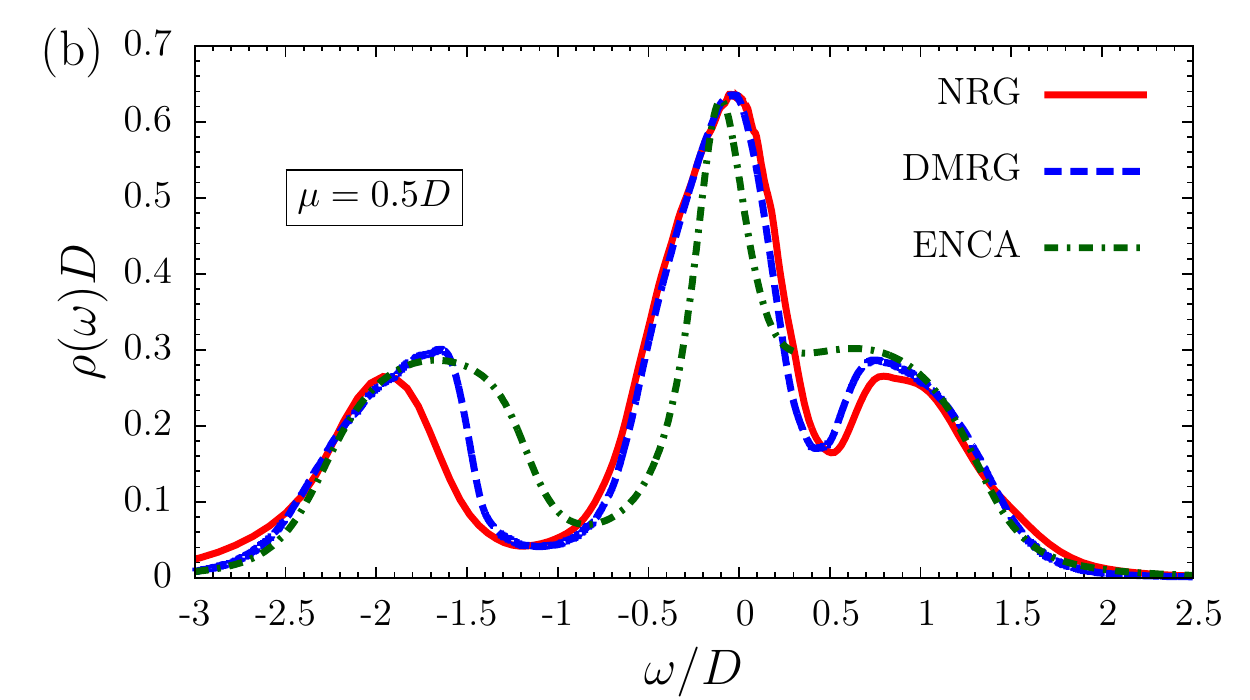}  
  \caption{(Color online)
    Spectral densities at $U=2D$ and chemical potentials $\mu=0$ (panel (a)) 
    	and $\mu=0.5D$ (panel (b)).
    For NRG and DMRG $T=0$ holds while for the
    ENCA  $T=0.027 D\approx 0.15 T^*$ ($\mu=0$) and  
    $T=0.1D\approx 0.6 T^*$ ($\mu=0.5D$).
    For $\mu=0.5D$,  the fillings are $n_{\mathrm{NRG}}\approx 0.580$, 
    $n_{\mathrm{DMRG}}\approx 0.592$, and $n_{\mathrm{ENCA}}=0.567$.}
  \label{fig:gf_u2_muRun}
\end{figure}

Increasing the Coulomb repulsion to $U=2D$ strongly enhances the correlations 
in the system. At half-filling, this increase drives the system towards the MIT.
In the spectral densities depicted in Fig.~\ref{fig:gf_u2_muRun} for
half-filling ($\mu=0$) and for finite chemical potential
($\mu=0.5D$), the Hubbard bands are now well separated from the
many-body resonance at the Fermi level. The inner edges of the Hubbard bands
are rather sharp with slight peaks associated presumably with bound trions of
a quasiparticle and a particle-hole pair.\cite{Karski2005,Karski2008} 
In the ENCA spectra, such peaks have also been 
obtained,\cite{pruschkeTransportHM93}
but they are not observed here due to the relatively high temperature.
The spectra obtained by DMRG 
and by ENCA  are almost indistinguishable at high energies, e.g., for 
$|\omega+\mu|\gtrsim 1.5D$.
The NRG curve falls off slower for large energies 
due to the limited resolution at large energies mentioned previously.

The deviation between the ENCA and the RG results are partly due
to the finite temperature to be used in the ENCA evaluation. Another
part is due to a too low Fermi liquid scale $T^*$ procured by the ENCA. 
This will be shown in the following.

\begin{figure}[ht]
  \centering
   \includegraphics[width=\columnwidth]{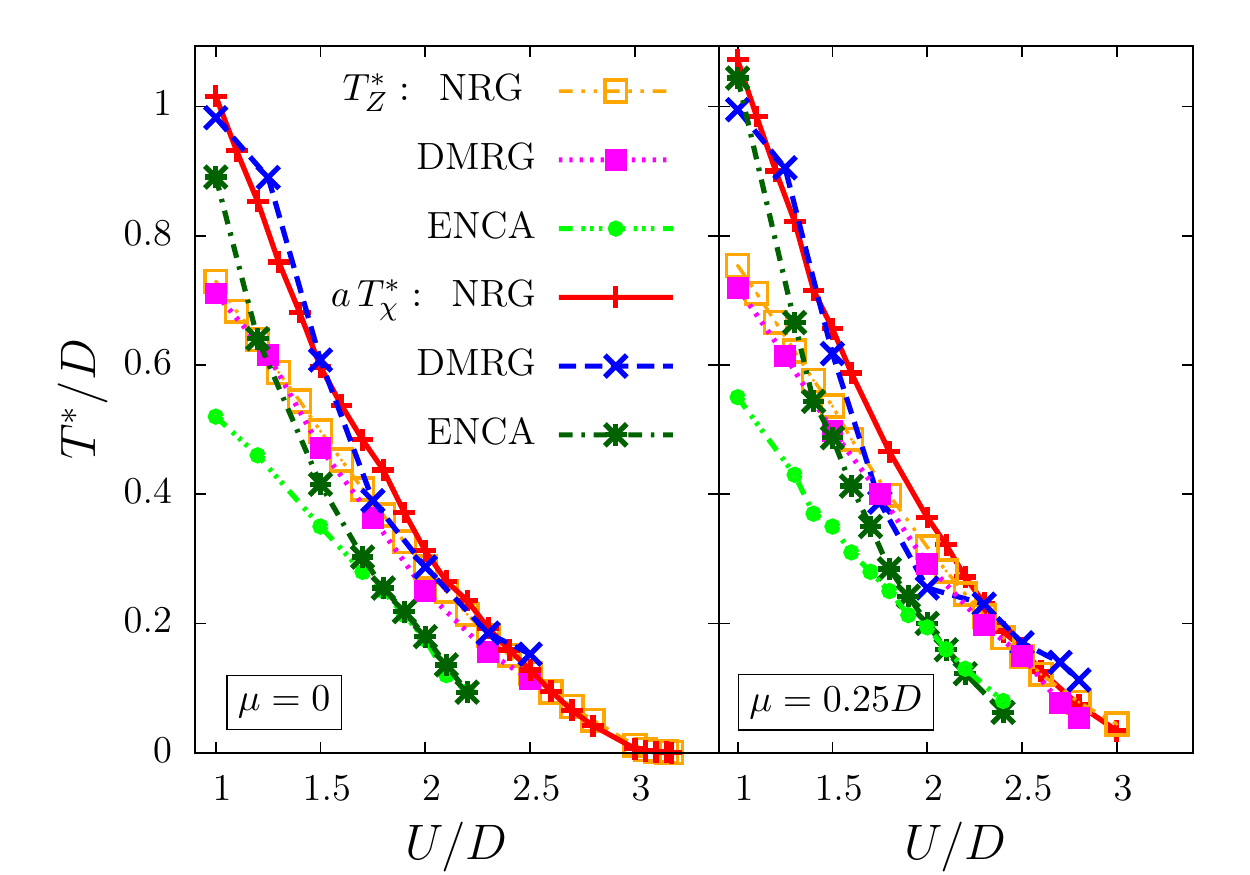}
  \caption{(Color online)
    Fermi liquid scales $T^*_Z$ and $T^*_\chi$ extracted 
    from the quasiparticle weight and from the maximum in the 
    spin susceptibility, respectively, as function of $U$ for two values of 
    the chemical potential. The magnetic scale $T^*_\chi$ is 
    rescaled by a \textit{single} factor $T^*_Z=a T^*_\chi=3.125\cdot T^*_\chi$ 
    to obtain coinciding energies.}
  \label{fig:scale}
\end{figure}

For further analysis, we extract a common energy scale
$T^*$ from the data of all three methods. 
In microscopic Fermi liquid theory  the low energy scale 
is proportional to the inverse mass enhancement\cite{varmaSingularFL02}
determined from the  quasiparticle weight $Z$.
The energy scale $T^*_Z$ defined in this way reads
\begin{equation}
  \label{eqn:qp-Z}
  T_Z^*=Z\,D=\frac{1}{1-\partial_\omega\mathrm{Re}\Sigma(0)}D
\end{equation}
and it is shown in Fig.~\ref{fig:scale} as function of the Coulomb repulsion 
$U$. For the two values of $\mu$ displayed in the figure,  $T^*_Z$ diminishes 
with increasing $U$   and vanishes at the MIT.
For larger values of the  chemical potential, e.g., 
$|\frac U2\pm\mu|\gtrsim\frac D2$,
the MIT will not occur anymore due to finite doping
and $T^*_Z$ will approach a constant value (not shown).

While the energy scale $T^*_Z$ extracted from the two RG approaches agree 
quantitatively, the ENCA scale follows the same trend, but is lower by about 
a factor of two. This in accord with results for the SIAM, i.e., without 
self-consistency, where the ENCA is known to produce the correct order of 
magnitude for the Kondo scale.\cite{pruschkeENCA89,schmittSus09}
The ENCA provides the exponential dependence of $T_\mathrm{K}$
on $U$, but the absolute values are slightly too low.\cite{greweCA108} 
Within the self-consistency of the DMFT this tendency persists and it is 
slightly amplified.

Another estimate for the low energy scale can be 
defined from the characteristic excitation energies for spin 
fluctuations.\cite{jarrell:dynamicSusSIAM91}
We determine a magnetic scale $T^*_\chi$ from
the position of the maximum in the local dynamic magnetic susceptibility,  
$T^*_\chi =\omega_{\mathrm{max}}$.
This estimate is equivalent to the energy scale extracted from the linear
slope in $\mathrm{Im}\chi_\mathrm{mag}(\omega)$ for small $\omega$.
As can be seen in Fig.~\ref{fig:scale},  the two scales $T^*_Z$ and $T^*_\chi$ 
of each method lie on top of each other at large $U$
if $T_\chi^*$  is rescaled by a \textit{single} factor $a$ of order unity.
Thus the low energy magnetic excitations and the 
single-particle excitations originate from the same physical process
which is governed by a single energy scale. We will call
such a behavior `universal' in the context of the present work.
For  the SIAM such behavior is well known to occur in the Kondo regime.
The DMFT self-consistency alters only quantitative aspects, 
but no qualitative ones.
Hence Fig.\ \ref{fig:scale} indicates universality in the metallic 
phase of the Hubbard model at large $U$
where Kondo-correlated quasiparticles dominate the low energy physics.  

We observe in Fig.~\ref{fig:scale} that the magnetic
scale $T^*_\chi$ and the single-particle scale $T^*_Z$
differ for small values of $U$ (and at large doping, not shown)
in analogy to what has been found in the SIAM.\cite{schmittAnders10}
$T^*_Z$ and $T^*_\chi$ differ so that no universality can be established.
A description of all excitations in terms of 
a \textit{single} energy scale cannot be maintained.
The Fermi liquid description is certainly still applicable, but all 
Landau parameters have to be determined independently.

Henceforth, we write $T^*=T^*_Z$ to represent the low energy 
scale and omit subscripts for simplicity.
 
\begin{figure}[ht]
  \centering
  \includegraphics[width=\columnwidth]{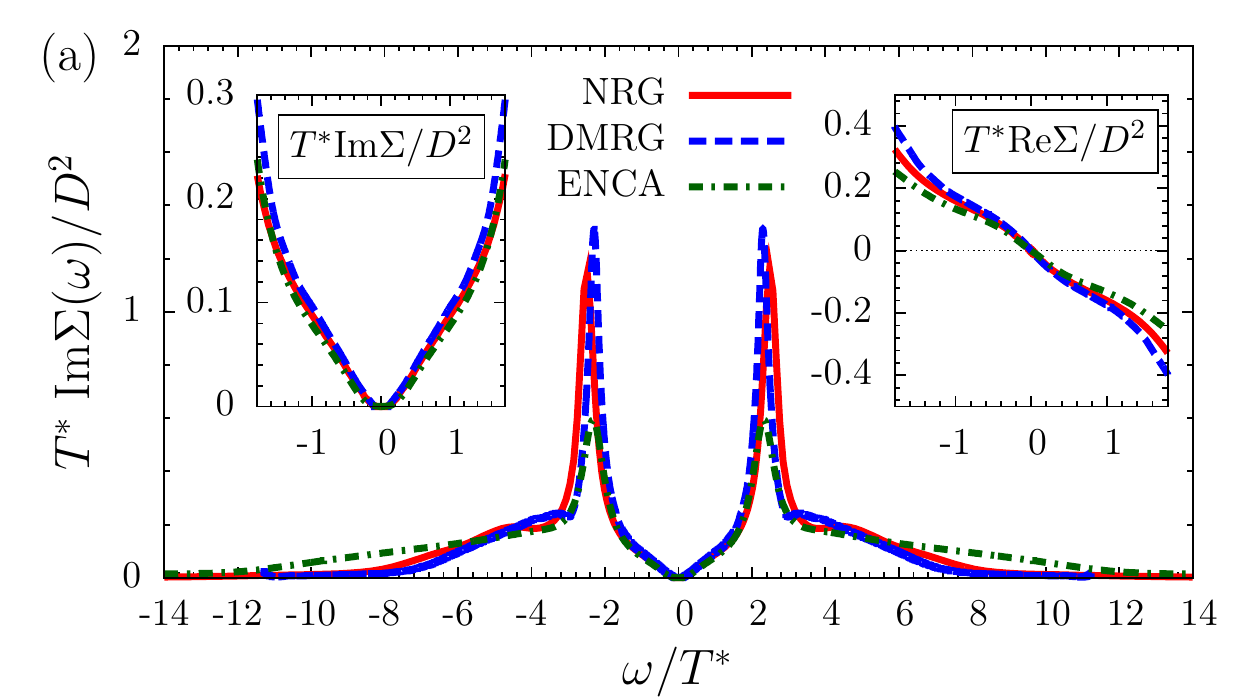}\\
  \includegraphics[width=\columnwidth]{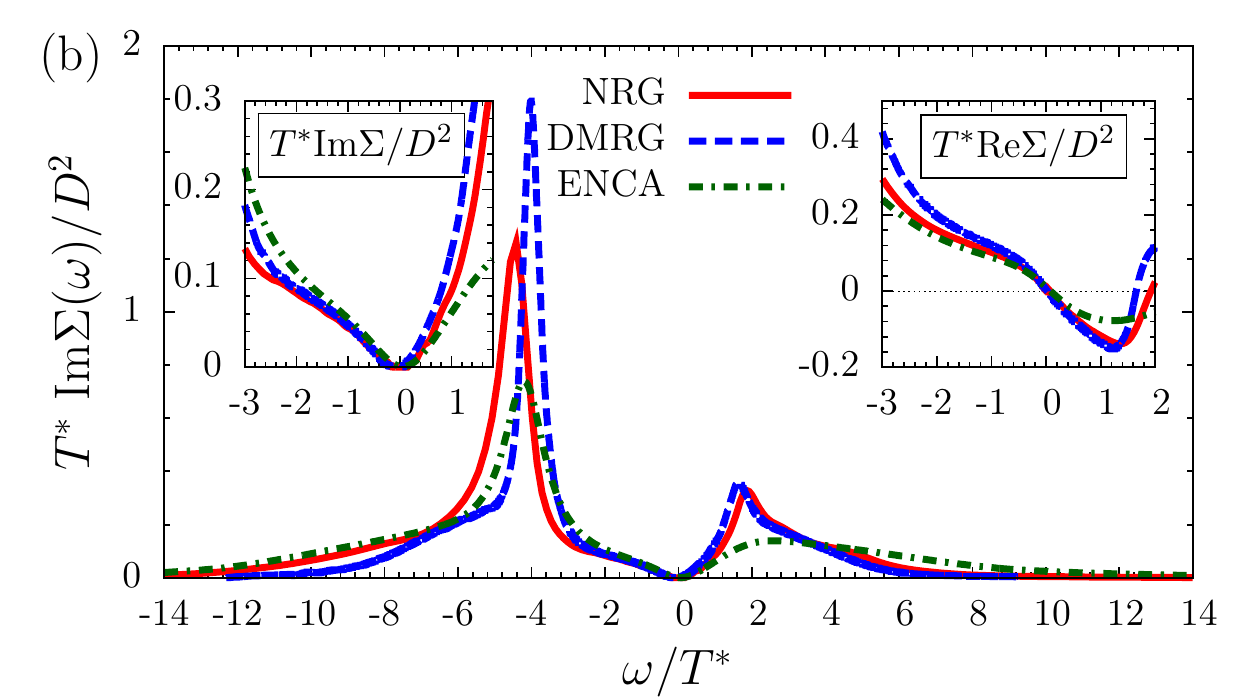}
  \caption{(Color online)
    Rescaled self-energy $T^* \Sigma(\omega)$ for $U=2D$ and (a) $\mu=0$ (b) 
    $\mu=0.5D$ as a function of energy measured in units  
    of the low energy scale $T^*$. Within the RG approaches we find 
    $T^*\approx 0.31D$ and        $T^*\approx 0.33D$ for $\mu=0$ and 
    $\mu=0.5D$, respectively. In ENCA we find 
    $T^*\approx 	0.18D$ for and $T^*\approx 0.2D$ for $\mu=0$ and 
    $\mu=0.5D$, respectively.
    The temperatures are as in Fig.~\ref{fig:gf_u2_muRun}.
    }
   \label{fig:sig_u2_muRun}
\end{figure}

The rescaled self-energy $T^* \Sigma(\omega)$, which determines the 
scattering rate of the Fermi liquid,\cite{varmaSingularFL02}  
is plotted in Fig.~\ref{fig:sig_u2_muRun} as function of $\omega/T^*$
for the same parameters as in Fig.~\ref{fig:gf_u2_muRun}.
Generally, the agreement between the three methods is very good and 
deviations only occur at large energies, establishing the too low
energy scale to be the main source of discrepancy between the ENCA
and RG methods. The  deviations for $\omega/T^*\gtrsim 1$ observable 
in panel (b) 
are on the one hand due to the improper description 
of correlated valence fluctuations\cite{schmittPhD08,schmittSus09}
and on the other hand due to the thermal broadening required in ENCA.

\begin{figure}[htb]
  \centering
  \includegraphics[width=\columnwidth]{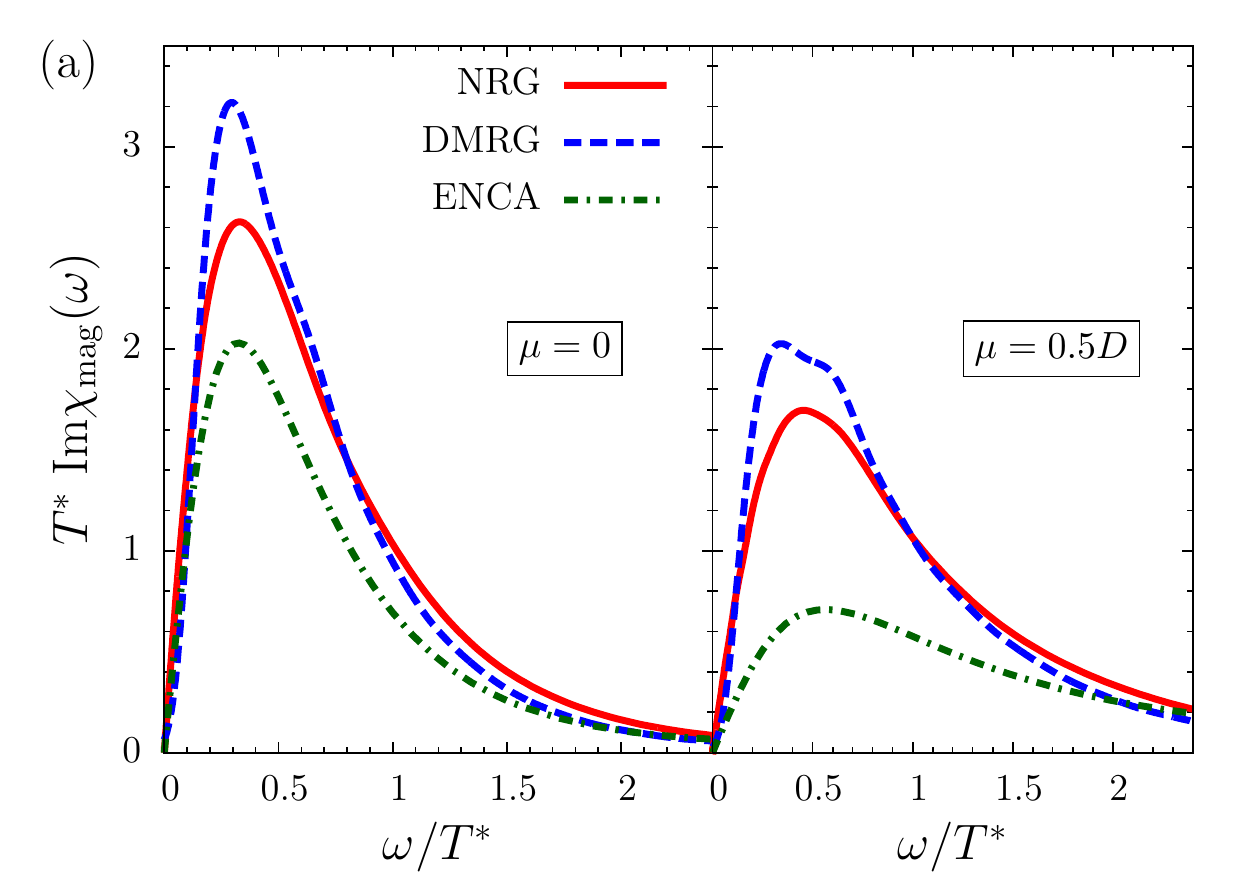}\\
  \includegraphics[width=\columnwidth]{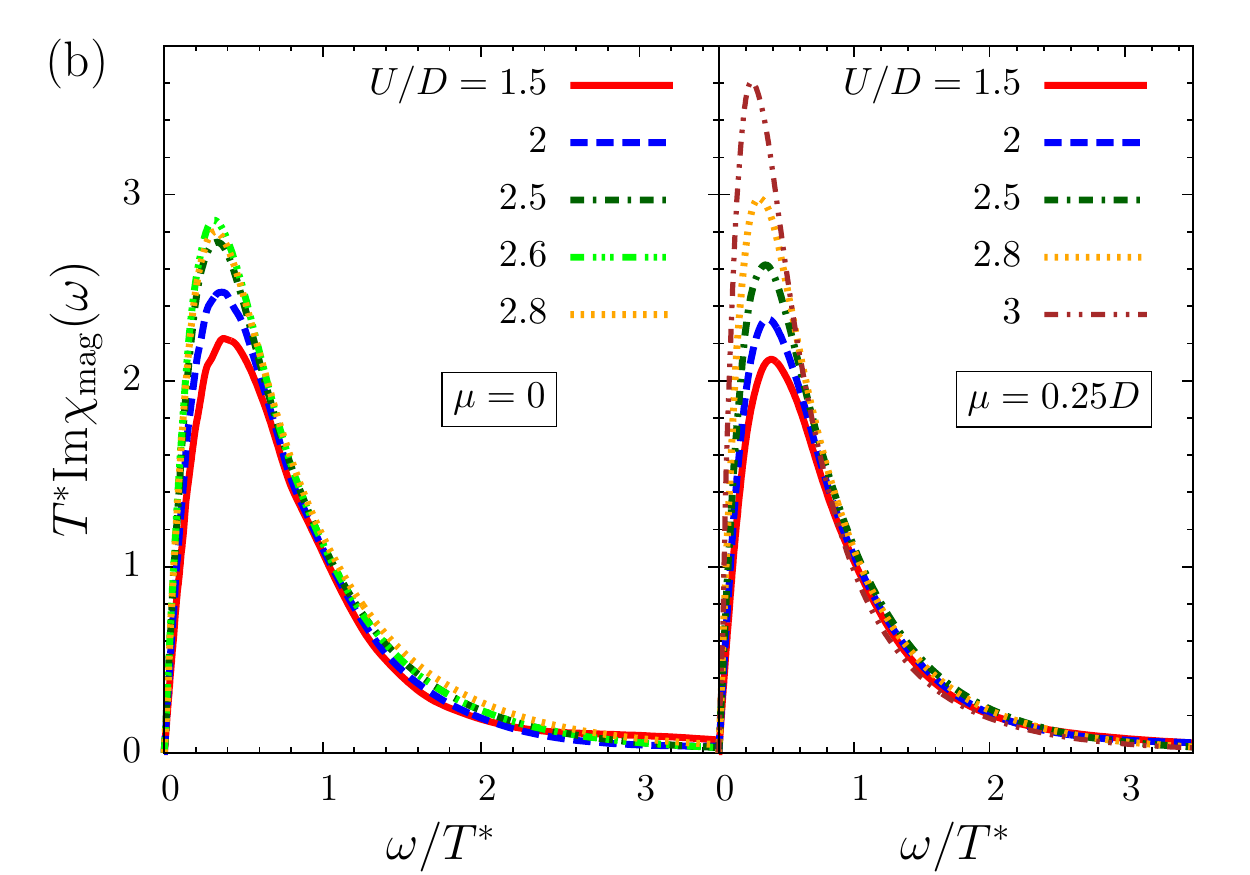}
  \caption{(Color online)
    (a) Rescaled imaginary part of the magnetic susceptibility 
    $T^*\mathrm{Im}\chi(\omega/T^*)$ for $\mu=0$ (left panel) and
    $\mu=0.5D$ (right panel) at $U=2D$ as a function of frequency. 
    Parameters are as in Figs.~\ref{fig:gf_u2_muRun}. 
    (b) $T^*\mathrm{Im}\chi(\omega/T^*)$ 
    vs.\ $\omega/T^*$  calculated with DMFT(NRG)
    for $\mu=0$ (left panel) and $\mu=0.25D$ (right panel) at various
    values of $U$.}
  \label{fig:sus_mag_u2}
\end{figure}

The rescaled dynamic magnetic susceptibility is depicted in 
Fig.~\ref{fig:sus_mag_u2}  for two different chemical potentials.
The peak positions from all three methods coincide but their heights differ.
For the ENCA this is due to the finite temperature $T\propto\mathcal O (T^*)$
and the susceptibility is expected to increase if $T\to0$.  
The use of raw NRG data without using the  equation-of-motion 
trick\cite{bulla:NRGSIAM98}
might be responsible for the discrepancies between the NRG 
and DMRG susceptibilities.

In the metallic regime, the universality conjectured before
is supported by the fact that $T^*\mathrm{Im}\chi_\mathrm{mag}(\omega/T^*)$ 
approaches a universal
function for large values of $U$  in the half-filled case. This can be 
observed  in Fig.~\ref{fig:sus_mag_u2}b where the susceptibility from NRG
is depicted for various values of $U$.
For finite $\mu$, however, $T^*\mathrm{Im}\chi_\mathrm{mag}(\omega/T^*)$ continues
to grow with decreasing $T^*$ indicating that valence fluctuations modify
the low energy physics decisively, thus abolishing universality in the lattice 
model away from half-filling. This is in contrast to what is found in the 
impurity model.

\begin{figure}[htb]
  \centering
  \includegraphics[width=\columnwidth]{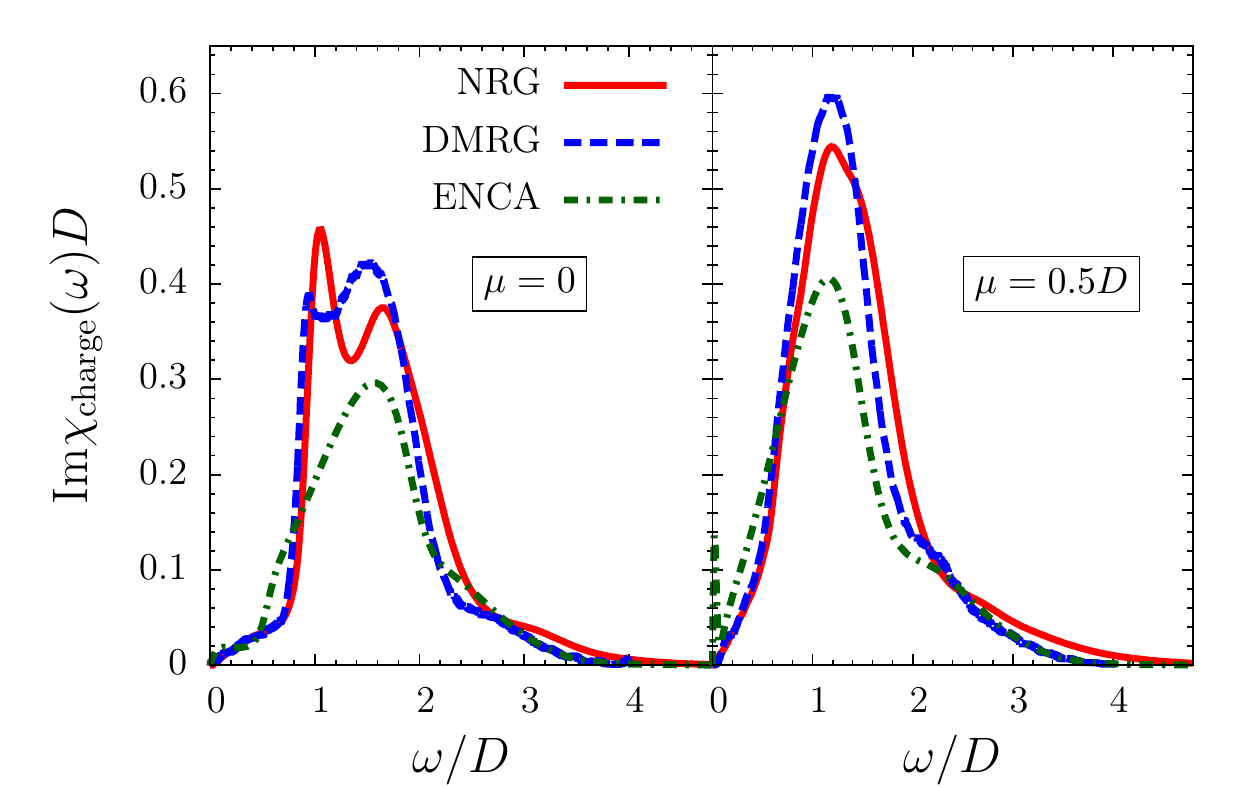}
  \caption{(Color online)
    Imaginary part of the charge susceptibility for $\mu=0$ (left panel) and
    $\mu=0.5D$ (right panel) at $U=2D$ as function of frequency in units of the
    half bandwidth $D$. Parameters are as in Fig.~\ref{fig:gf_u2_muRun}.}
  \label{fig:sus_charge_u2}
\end{figure}

The local charge susceptibility $\mathrm{Im}\chi_\mathrm{charge}(\omega)$ shown 
in Fig.~\ref{fig:sus_charge_u2}  is strongly suppressed for small energies 
$\omega$  as consequence of the large Coulomb repulsion. This is particularly 
striking in comparison to the spin susceptibility. 
The characteristic energy scale of the charge susceptibility remains set by the
interband excitation energy between the quasi-particle band and the Hubbard
bands which is of the order $\sim \frac U 2-|\mu|$.

In the curves  obtained by the RG methods at half-filling, peaks
emerge at the onset of the interband excitations as can be seen in
Fig.~\ref{fig:sus_charge_u2} (left panel). They originate 
from the sharp features at the inner Hubbard band edges.\cite{Raas2009a}
As in the single particle spectrum, these are missing in the ENCA
curve due to thermal broadening. 
Away from half-filling,  the charge susceptibility increases due to the 
increased phase space at low energies. There, the ENCA reveals its limitations 
in the mixed-valence regime since at low temperature its threshold exponents
generate an additional low frequency peak for $\omega\to 0$ which is expected
to disappear if higher order vertex corrections were included.

For increasing Coulomb repulsion $U$, the spectral weight of 
$\mathrm{Im}\chi_\mathrm{charge}$ is shifted towards larger energies $\omega$ as
illustrated in Fig.~2 in Ref.\ \onlinecite{Raas2009a}. In contrast, 
the position of the maximum  of the spin susceptibility, which sets the scale 
$T^*_\chi$, is shifted towards  smaller energies with increasing $U$ as shown 
in Fig.~\ref{fig:sus_mag_u2}b.  As a consequence, the energy scales of 
collective charge and spin excitations are clearly separated. 
This will turn out to be important for the observation of kinks.

\begin{figure}[ht]
  \centering
   \includegraphics[trim=0.4cm 0cm 0cm 1.5cm,clip=true,width=9cm]{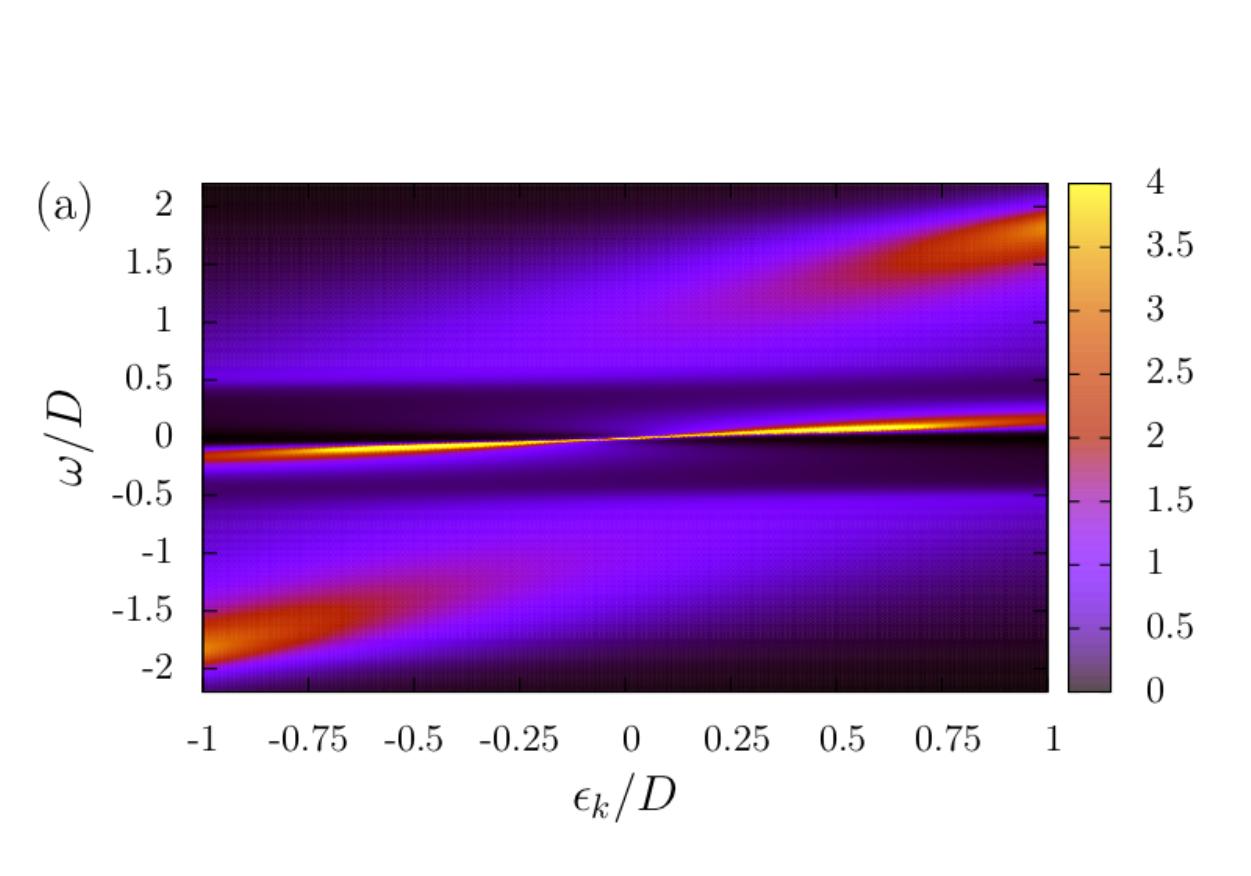}\\
  \includegraphics[trim=0.4cm 0cm 0cm 1.5cm,clip=true,width=9cm]{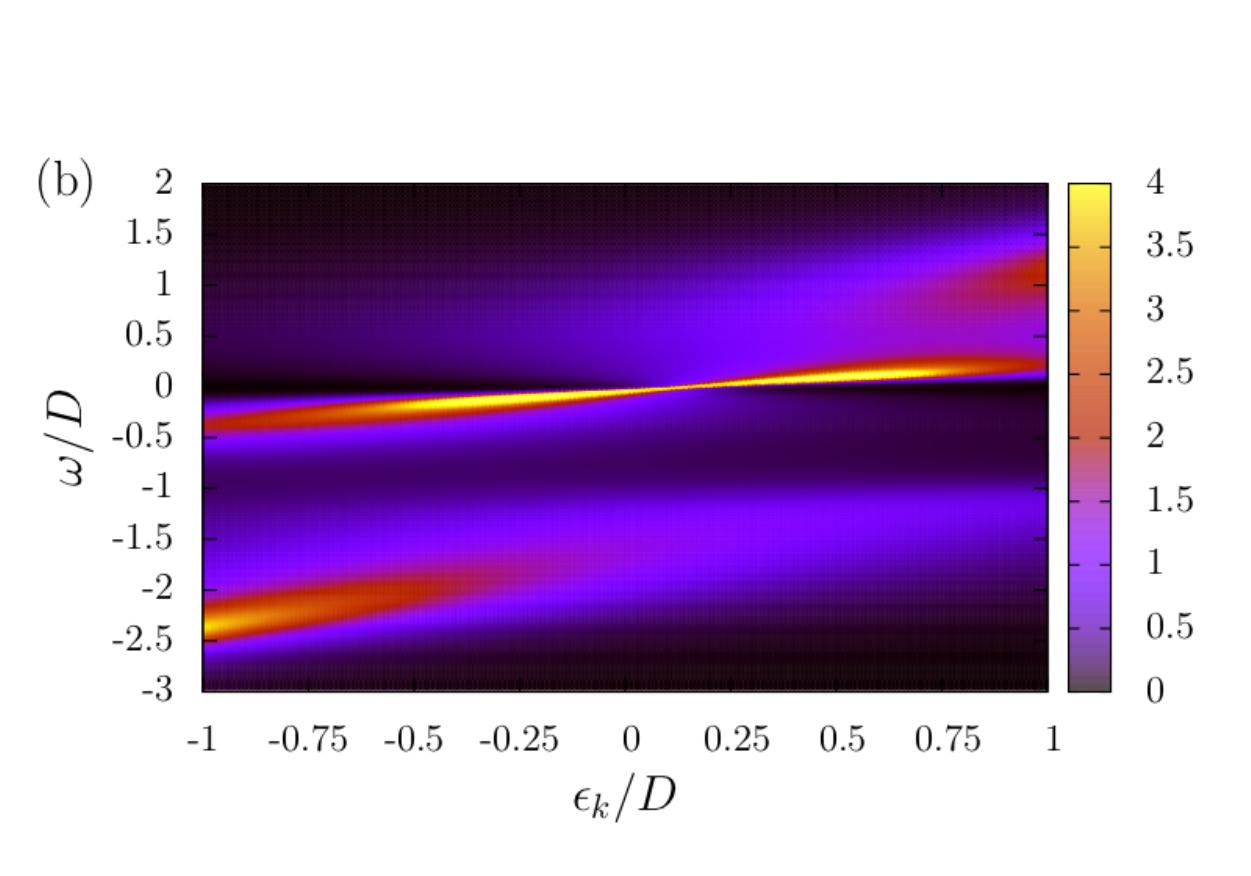} 
  \caption{(Color online)
    Spectral density $\rho(\omega,\epsilon_k)$ obtained within DMFT(ENCA) as
    function of frequency $\omega$ and bare electronic energy $\epsilon_k$ 
    for $U=2D$ at, (a) half-filling ($\mu=0$) and (b) $\mu=0.5D$.}
  \label{fig:gfk_rest}
\end{figure}

Finally, we illustrate the full dependence of the spectral densities
on the momentum via the bare dispersion $\epsilon_k$ according to
\begin{equation}
  \label{eq:gfk}
  \rho(\omega,\epsilon_k) =
  \frac{1}{\pi}\mathrm{Im}\frac{1}
       {\omega-\mathrm{i}\delta-\epsilon_k-\Sigma(\omega)}
\end{equation}
in the false-color plots in Fig.\ \ref{fig:gfk_rest}.
The separation of single-particle energy scales and the influence of
particle-hole asymmetry can be seen clearly. The almost
flat ridge around $\omega=0$ represents the narrow band of
heavy quasiparticles which is well-separated in energy from the lower and 
upper Hubbard bands below and above $\omega=0$. The coherence of the 
quasiparticle excitations is lost once  $\epsilon_k$  reaches the scale 
$T^*_\chi$ where spin fluctuations become
important. This causes a kink in $\mathrm{Re}\Sigma$ as will be discussed in 
the following section. At large electron doping shown Fig.~\ref{fig:gfk_rest}b,
the upper Hubbard band and the quasiparticle band merge rendering charge and 
spin excitations equally important for positive energies. However, at negative 
energies, valence fluctuations are suppressed and the separation of energy 
scales persists.

\section{Kinks and Collective Modes}
\label{sec:kinks-collective-modes}

\label{sec:kinks}

It is well known that the coupling of fermions to energetically
low-lying bosonic modes causes kink-like structures in the fermionic dispersion.
This picture is based on diagrammatic weak fermion-phonon coupling theory,
see for instance Refs.\ \onlinecite{engel63} and \onlinecite{scala69}, and 
references therein.
The kink in the fermionic dispersion occurs roughly at the bare phonon
energy. For stronger coupling, the diagrammatic approach breaks down 
\cite{bened98,meyer02} and the kink feature persists but it does
no longer occur at the bare phonon energy.\cite{bauer10}
Roughly, the strength of the kink increases with the coupling between
the fermionic and the bosonic modes.

Recently, it was demonstrated that kinks in the electronic dispersion are a 
generic feature of strongly correlated electron systems without any
coupling to external bosons.\cite{byczukKinks09}
Subsequently, it was shown that the kinks occurring in strongly
interacting electron systems can be seen as result of the coupling
of the fermions to the emergent collective excitations of magnetic character.
Thus the system creates its own bosonic modes which in return generate
the kinks.\cite{Raas2009}

The original argument by Byczuk \textit{et al.} for the kinks was based on the 
three-peak structure in the spectral density $\rho(\omega)$,
as shown in Fig.~\ref{fig:gf_u2_muRun}. Its essence is as follows:
The many-body resonance extends around the Fermi level $\omega=0$ from
$\Omega_{-}<0$ to $\Omega_{+}>0$, where $\Omega_\pm$ are the positions
of the minima between the many-body resonance and the Hubbard bands.

\begin{figure}[htb]
  \centering
  \includegraphics[width=\columnwidth]{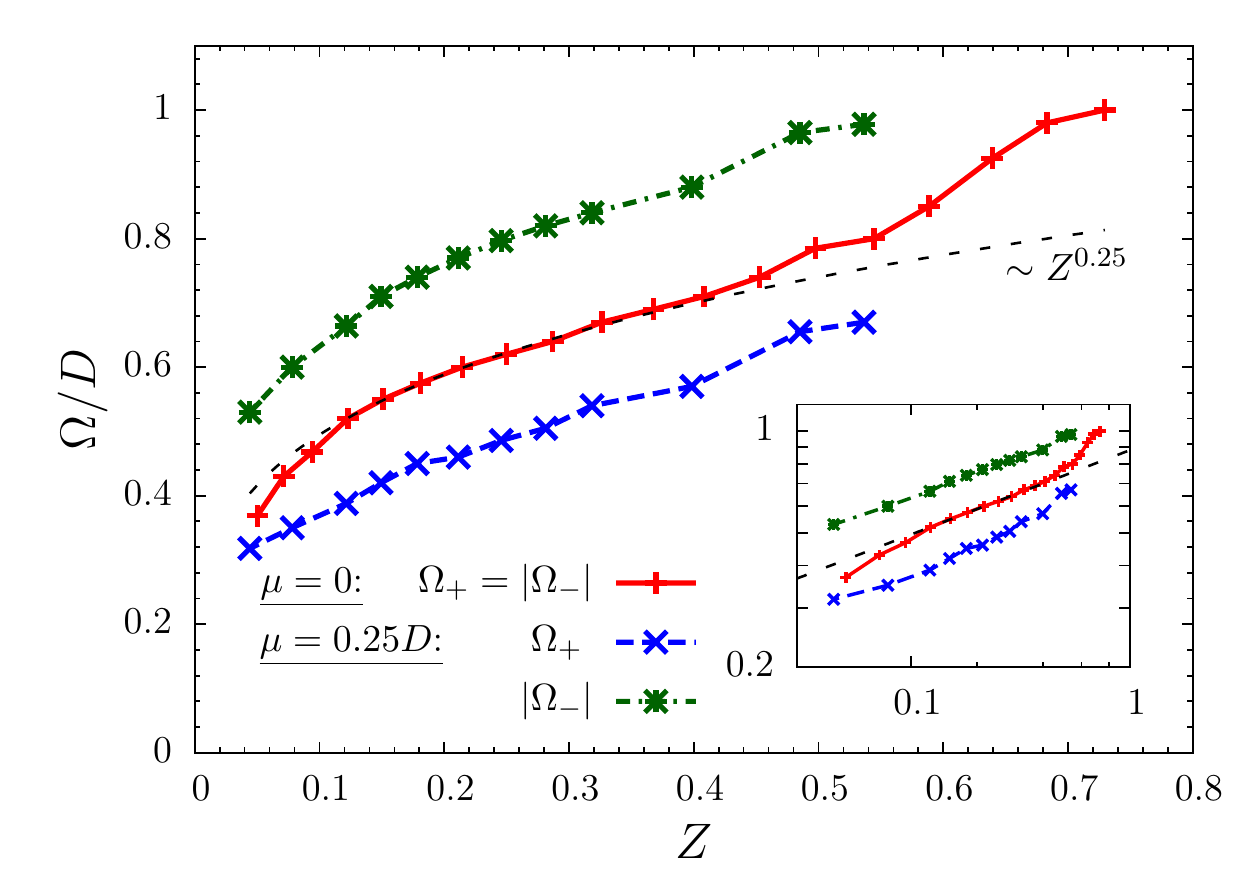} 	
  \caption{(Color online)
    Positions of the minima between the many-body resonance and the Hubbard
    bands calculated by NRG as function of the quasiparticle weight
    $Z$. The inset shows the same data in a double-logarithmic plot. 
    The dashed line depicts a power law $Z^{0.25}$ for comparison.}
  \label{fig:OmegaZ}
\end{figure}

Numerical results for the positions $|\Omega_\pm|$ of these minima 
calculated by NRG are shown in Fig.~\ref{fig:OmegaZ} as 
function of the quasiparticle weight $Z$. The lifting of the degeneracy of 
$|\Omega_{\pm}|$ upon doping is clearly visible. The NRG analysis suggests that 
the power law scaling $\Omega\propto Z^{1/4}$ holds for $Z\to 0$ (dashed line 
in Fig.~\ref{fig:OmegaZ} and in its inset).

Furthermore, the KK relation implies that the real part of $G(\omega)$ is 
linear around  $\omega=0$ and that it has sharp maxima at positions $r_\pm$ 
which are of the order of the half-width of the many-body resonance.
In particular $|r_\pm |<|\Omega_\pm|$ holds.
For the semi-elliptic DOS the self-consistency condition for the effective 
medium can be solved analytically by $\Gamma(z)=(D/2)^2G(z)$. 
For general lattices, using this relation amounts up
to approximating  $\Gamma[G(z)]$ by the linear term of its 
moment expansion. Then Eq.~\eqref{eq:dmftSig}
implies kinks in $\mathrm{Re}\Sigma(\omega)$
at $\omega_{\star}^\pm$ with $|\omega_{\star}^\pm|<|\Omega_\pm|$.

An explicit expression for the position of the kinks can
be obtained by describing the local Green function around the Fermi level by
\begin{align}
  G(\omega)& \approx \frac{Z_{\text{loc}}}{\omega-\omega_0-
    \mathrm{i}(\gamma+\gamma'\omega)}.
\end{align}
The parameters $\omega_0$, $\gamma$, $\gamma'$, and $Z_{\text{loc}}$
are determined from the physical quasiparticle weight $Z$ 
and the non-interacting DOS alone. The kink positions $\omega_{\star}^{\pm}$
are calculated as the maxima of the second derivative of
the dressed dispersion
\begin{equation}
  \label{kink-formula} 
  \omega_{\star}^{\pm} = 
  \omega_0\mp\frac{\gamma+\gamma'\omega_0}{\sqrt{1+\gamma'^2}}
  \left(1-\sqrt{2}\sqrt{1\mp\frac{\gamma'}{\sqrt{1+\gamma'^2}}}\right),
\end{equation}
for details see supplement of Ref.\ \onlinecite{byczukKinks09}, but note the 
differing sign of the inner square root. For the particle-hole symmetric case 
one has  $\gamma'=\omega_0=0$  so that \eqref{kink-formula} reduces to 
$\omega_{\star}^{\pm}=\pm(\sqrt{2}-1)\gamma$. For the semi-elliptic
DOS $\gamma=ZD$ the kinks are thus located at 
$\omega_{\star}^{\pm}=\pm(\sqrt{2}-1)ZD$.

\begin{figure}[ht]
  \centering
  \includegraphics[width=\columnwidth]{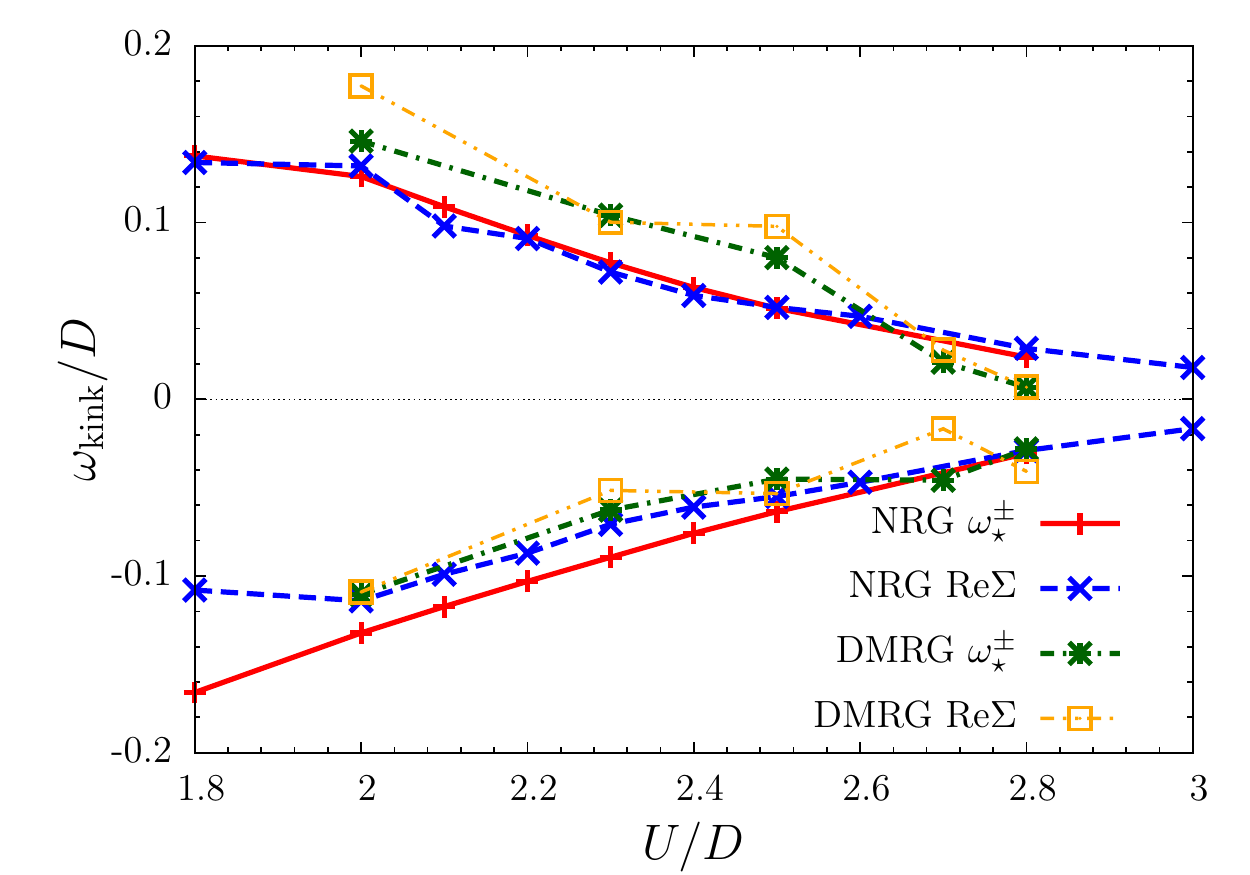}
  \caption{(Color online) 
    Comparison of the kink position from a numerical analyis of 
    $\mathrm{Re}\Sigma$ and
    from Eq.~(\ref{kink-formula}) for $\mu=0.25D$ for the two RG impurity
    solvers as function of  $U$.}
  \label{fig:kinkformcheck}
\end{figure}

Indeed, the kink positions in the numerical data for $\mathrm{Re}\Sigma$
agree nicely with results obtained via Eq.~(\ref{kink-formula}). This is
demonstrated for $\mu=0.25D$ in Fig.~\ref{fig:kinkformcheck} where the kink
positions extracted for the DMRG and the NRG impurity solver 
are plotted as function of $U/D$.
Only data for small doping and large repulsion is shown because for large doping
and/or small repulsion  the three-peak structure of $\rho(\omega)$ is not found
so that the above analytical argument does not hold
and the kinks cannot be  resolved. 
For large doping and strong repulsion there still exists one kink in 
$\mathrm{Re}\Sigma$ for which
Eq.~(\ref{kink-formula}) still predicts the correct position (not shown).

The above analysis due to Byczuk et al.\ focusses on the mathematical structure
of the equations defining the propagator and the self-energy. In Ref.\ 
\onlinecite{Raas2009} some of us showed for the half-filled case that the 
characteristic excitation energy of spin fluctuations agrees with the kink 
energy.  Kinks in $\mathrm{Re}\Sigma(\omega)$ lead via the KK relation to 
inflexion points in $\mathrm{Im}\Sigma(\omega)$ at the same energies. 
This corresponds to a change in the quasiparticle lifetime 
$\tau \sim 1/\mathrm{Im}\Sigma$. Inversely, this implies that humps in 
$\mathrm{Im}\Sigma$ imply kinks in $\mathrm{Re}\Sigma(\omega)$
via the KK transform. Hence, even though  no explicit bosonic
modes are included in the Hubbard model, the emergent collective spin 
excitations are responsible for the structures in $\mathrm{Im}\Sigma(\omega)$ 
and thus for  the kinks in $\mathrm{Re}\Sigma$.

\begin{figure}[htb]
  \centering
  \includegraphics[width=\columnwidth]{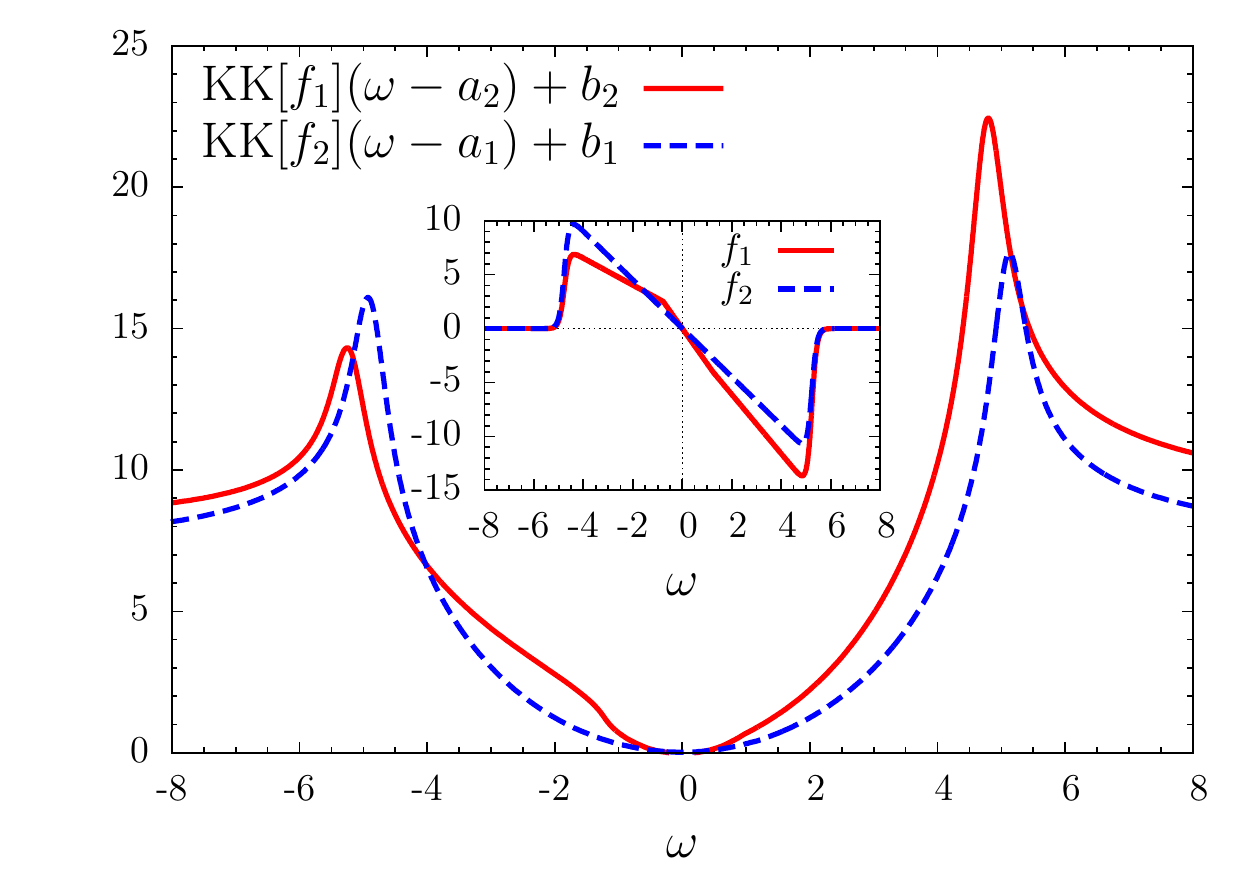}
  \caption{(Color online)
     Illustration of the KK transforms of two test functions $f_1$
     	and $f_2$, which are depicted in the inset. The first, $f_1$ displays
     	a kink while the second ($f_2$) does not, but has the same average slope
     	for small frequencies. The KK transforms are shifted by $a_i, b_i$ such
     	that KK$[f_i](-a_i)+b_i=0$.
  }
  \label{fig:asymmetric_illu}
\end{figure}

\begin{figure}[htb]
  \centering
  \includegraphics[width=\columnwidth]{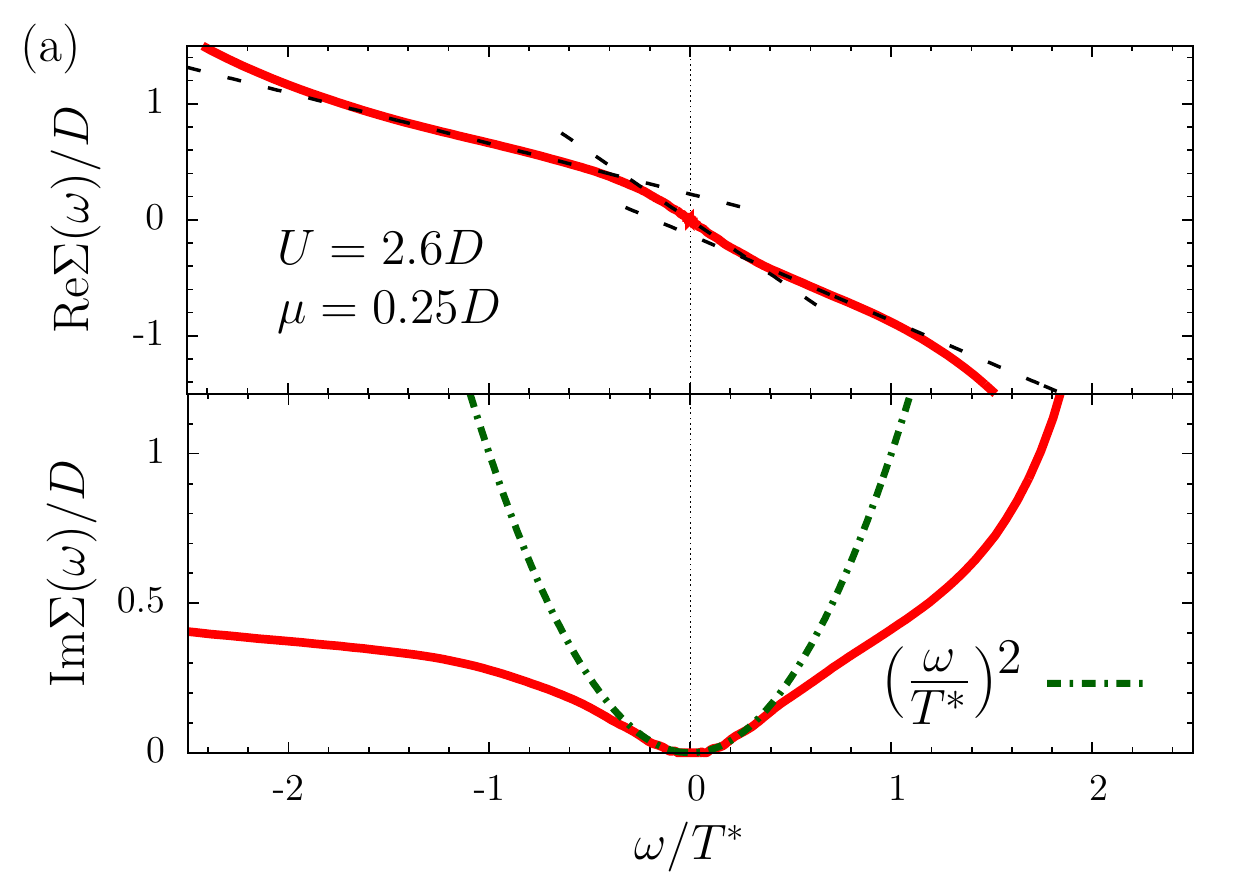}\\
  \includegraphics[width=\columnwidth]{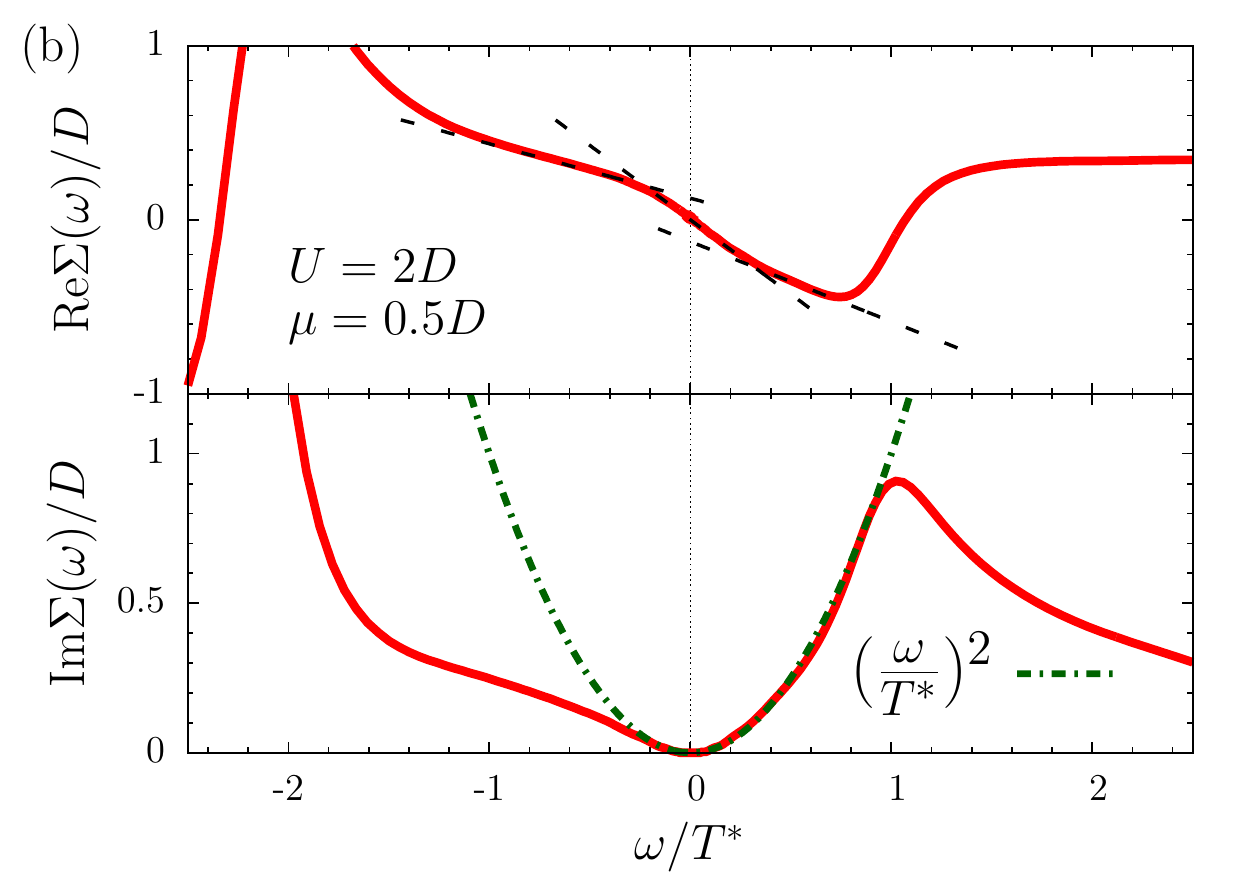}		
  \caption{(Color online)
    (a) $\mathrm{Re}\Sigma(\omega)$ (upper panel) and 
    $\mathrm{Im}\Sigma(\omega)$
    (lower panel) at finite doping as function of $\omega/T^*$ obtained from
    DMFT(NRG). The dashed lines in the upper panel indicate linear fits 
    used to determine kink positions. In the lower panel 
    the parabola $(\omega/T^*)^2$ expected from Fermi liquid theory is included
    for comparison. The coherence scale $T^*\approx 0.16 D$ is the low-energy 
    scale for $U=2.6D$, cf.\ Fig.~\ref{fig:scale}.
    (b) Same as in panel (b), but for $\mu=0.5D$ at $U=2D$. Note that the 
    kink at positive of frequencies is very weak and concomitantly 
    $\mathrm{Im}\Sigma(\omega)$ displays parabolic behavior 
    up to $\omega\approx T^*$ except for a very shallow hump.
  }
  \label{fig:asymmetric_real}
\end{figure}

This argument is also applicable without particle-hole symmetry, i.e.,
for the doped model. The kinks in $\mathrm{Re}\Sigma(\omega)$
are still associated with additional inflexion points in 
$\mathrm{Im}\Sigma(\omega)$ which are related to changes in the relaxation 
mechanism.  To illustrate  this view qualitatively, we mimic a 
kink in  $\mathrm{Re}\Sigma(\omega)$ by 
the function $f_1$ depicted in Fig.\ \ref{fig:asymmetric_illu} and
include for comparison $f_2$ without a kink. Then we study the 
differences in the KK transforms
which correspond to the imaginary part. While KK$[f_2]$ is governed by a wide
parabola in the range $\omega\in(-4,4)$, KK$[f_1]$ displays a noticeable hump 
starting below the frequency of the kink. A parabolic fit would hold only in 
the interval $\omega\in(-1,1)$.

DMFT  self-energies computed with the NRG are shown in Fig.\ 
\ref{fig:asymmetric_real}a. The dashed lines in the upper panel 
indicate the linear fits to $\mathrm{Re}\Sigma(\omega)$  used to determine the 
kink  positions. 
Fig.\ \ref{fig:asymmetric_real}a displays the same qualitative features as 
Fig.\ \ref{fig:asymmetric_illu}  though they are less pronounced.  
The physical model does not display mathematically sharp  kinks as the test 
function $f_1$ does.  The real part $\mathrm{Re}\Sigma(\omega)$ in Fig.\ 
\ref{fig:asymmetric_real}a displays
two kinks.  The one at negative frequencies is fairly clear, the one at positive
frequencies is fairly weak. Correspondingly, the humps in 
$\mathrm{Im}\Sigma(\omega)$ are clearly visible at negative frequencies, but 
only weakly discernible at positive frequencies. 
The kinks, which mark the beginning of the humps, indeed occur at about 
$T^*_\chi=\frac13T^*(\equiv \frac13T^*_Z)$, cf.\ Fig.\ 
\ref{fig:scale}, in agreement with the previous finding at 
half-filling.\cite{Raas2009}

A parabolic description in terms of the Fermi liquid scale $(\omega/T^*)^2$
is possible, but only up to about $|\omega/T^*|\approx 0.3$, again  
in accord with the  finding at half-filling.\cite{Raas2009}
The scattering rate as given by $\mathrm{Im}\Sigma(\omega)$ decreases 
compared to a pure $(\omega/T^*)^2$ behavior with increasing $|\omega|$.

This picture is consistent with the renormalization group  flow  and the RG 
fixed points  of the effective site for a converged  metallic DMFT solution. 
For $T^*\gg |\omega|\to 0$, the physics is determined by 
a line of strong coupling (SC) fixed  points which describes a Fermi-liquid 
with broken particle-hole 
symmetry.\cite{krishnamurtyNRGSIAMII80,bullaNRGReview08} Its
characteristic energy scale is given by $T^*$,  and  
$\mathrm{Im} \Sigma(\omega) \propto (\omega/T^*)^2$. 
With increasing frequency, however, the system 
is described by the unstable local-moment (LM) fixed point.
The dynamic Kondo singlet is broken on a scale $T_\mathrm{K}\propto T^*_\chi$ by
singlet-triplet excitations and the quasiparticles disintegrate 
at higher excitation energies, leaving a free local spin coupled to the 
conduction band.  As a consequence, the scattering is reduced and the 
self-energy is increasing much slower than close to the SC fixed point. 
Spin-flip scattering dominates the self-energy in the vicinity of the 
LM fixed point.

Therefore,  the single-particle self-energy 
retraces the crossover from the LM to the SC fixed point.
At very high frequencies the magnetic scattering is weak, and the physics of 
the Hubbard model is determined by the local charging energies derived from 
the atomic  picture. Hence the self-energies depicted in 
Fig.~\ref{fig:sig_u2_muRun} are low and featureless at very high frequencies.
At intermediate frequencies, the spin-flip scattering provides an additional 
decay channel on top of a weakly correlated,  particle-hole asymmetry 
conduction band. This additional decay channel reveals itself in 
$\mathrm{Im}\Sigma(\omega)$ as the humps  at intermediate energies, 
cf.\ Figs.\ \ref{fig:asymmetric_illu}  and \ref{fig:asymmetric_real}.
At low energy scales, the spin-flip scattering is replaced in 
$\mathrm{Im}\Sigma(\omega)$ by the Fermi-liquid  parabola determined by 
$T^*$ for $|\omega|\to 0$.

At  larger electron doping, only the lower Hubbard band is well separated from 
the quasiparticle band, see Fig.~\ref{fig:gf_u2_muRun}.
The particle-hole asymmetry and the correlated valence fluctuations
matter for positive excitation  energies.
The scales for spin and charge excitations 
are not well separated so that both channels contribute to the self-energy. 
The $\mathrm{Im}\Sigma(\omega)$ remains almost 
quadratic in $\omega$ for $\omega>0$ even on the scale of
$\omega\approx T^*$ as can be seen in Fig.~\ref{fig:asymmetric_real}b
for $\mu=0.5D$ at $U=2D$. Only a minute hump occurs in the quasiparticle decay 
rate at the spin excitation energy and consequently only a very weak
kink occurs at positive energies. Even though the spin-susceptibility
shows a pronounced maximum, the accessibility of low-energy charge fluctuations
for positive energies suppresses the kink, in accord with the two conditions 
stated in the Introduction.

At negative excitation energies, correlated low-energy valence 
fluctuations cannot be excited so that the charge energy scale is well 
separated from the coherence scale $T^*$. Thus the kink and the corresponding 
hump are distinct at negative frequencies.

\begin{figure}[ht]
  \centering
  \includegraphics[width=\columnwidth]{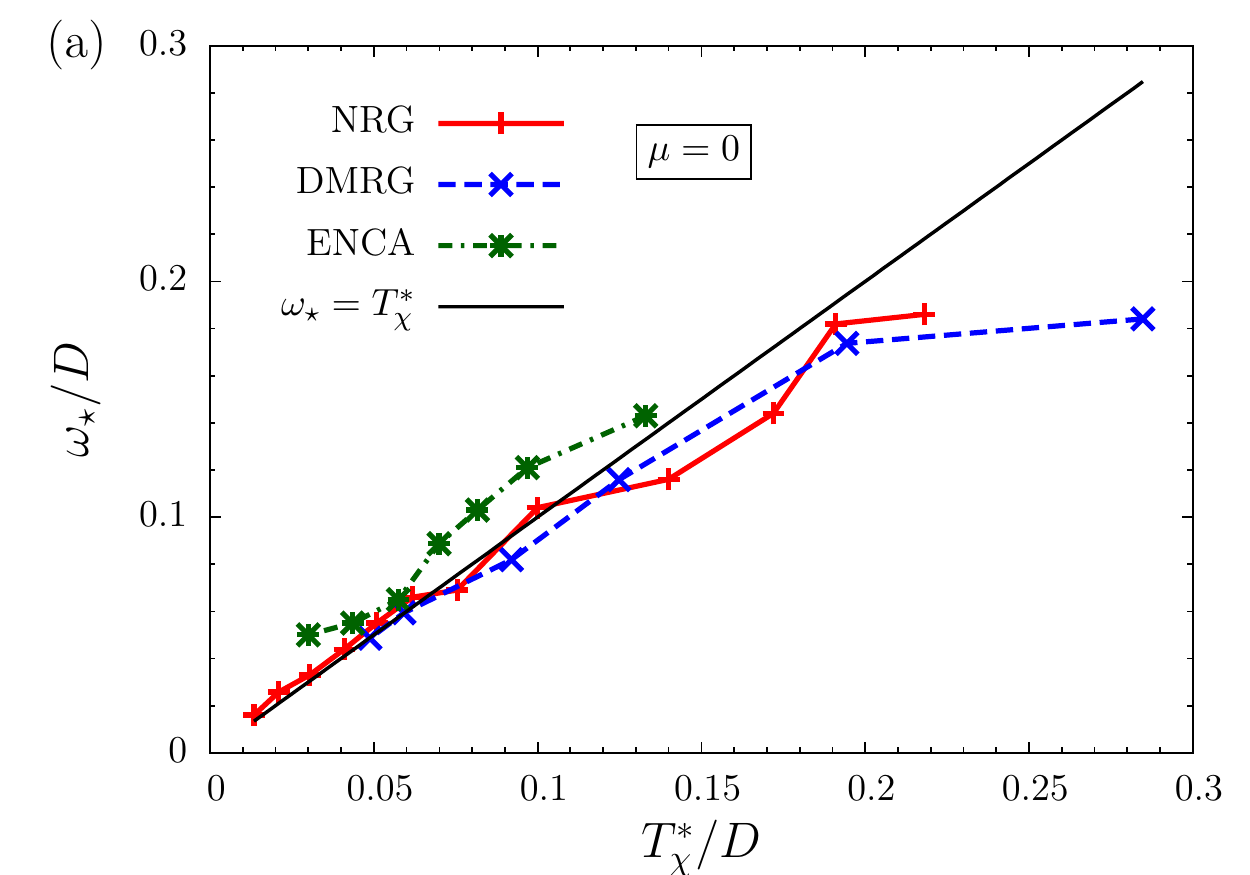}\\
  \includegraphics[width=\columnwidth]{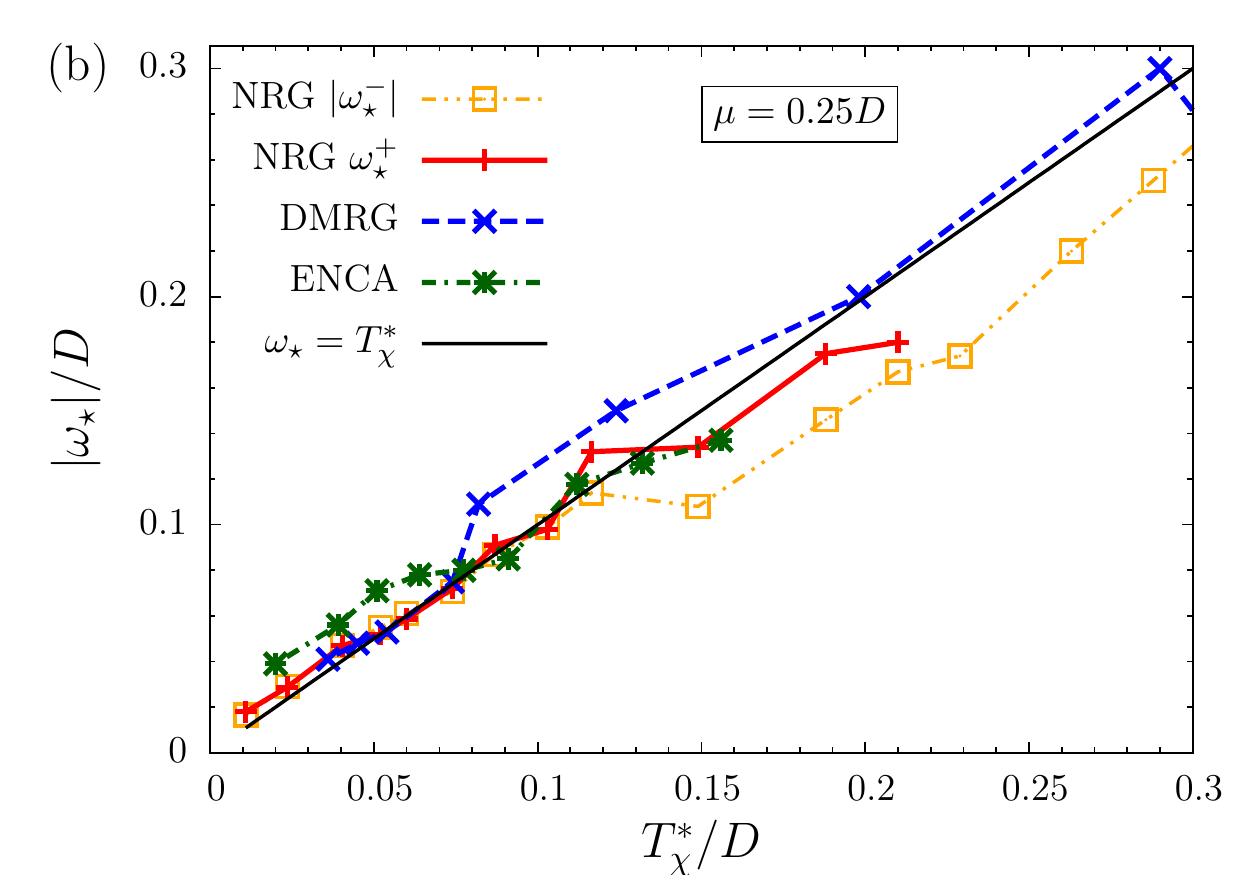} 
  \caption{(Color online)
    Kink energies $\omega_\star^\pm$ as function of the 
    frequency of the maximum $T^*_\chi/D$ in the imaginary part of the 
    spin susceptibility for (a) $\mu=0$ and (b) $\mu=0.25D$.
  }
  \label{fig:pos_kink_sus}
\end{figure}

The above discussion shows that 
kinks can be directly linked to physical processes in the system.
As in the symmetric case, the  kink positions in the self-energy correlate
with the positions of the maxima of the spin susceptibility, i.e., $T^*_\chi$,
which is shown in Fig.~\ref{fig:pos_kink_sus} for two values of $\mu$. 
The values for both quantities from all three methods coincide and 
the small deviations can be understood from the strengths and the weaknesses
of the methods as discussed in Sect.\ \ref{sec:comparison}.
The kink positions $\omega_\star$ equal  $T^*_\chi$ for $T^*_\chi\to 0$.
This clearly supports the view that the spin fluctuations are responsible for 
the kinks.  Deviations occur for larger energies corresponding to smaller 
values  of $U$. There, the Fermi liquid description in terms of a single energy
scale does not apply anymore since $T^*_\chi$ and $T^*_Z$ stem from combinations
of different types of excitations, and  non-universal valence fluctuations play
a role. Additionally, kinks are less pronounced for small $U$  and hence their 
positions are harder to determine unambiguously. For smaller $U$ the 
spin fluctuations do not yet behave like emergent collective bosonic modes, 
and the charge and spin excitations are not well separated in energy.

\begin{figure}[htb]
  \centering
  \includegraphics[width=\columnwidth]{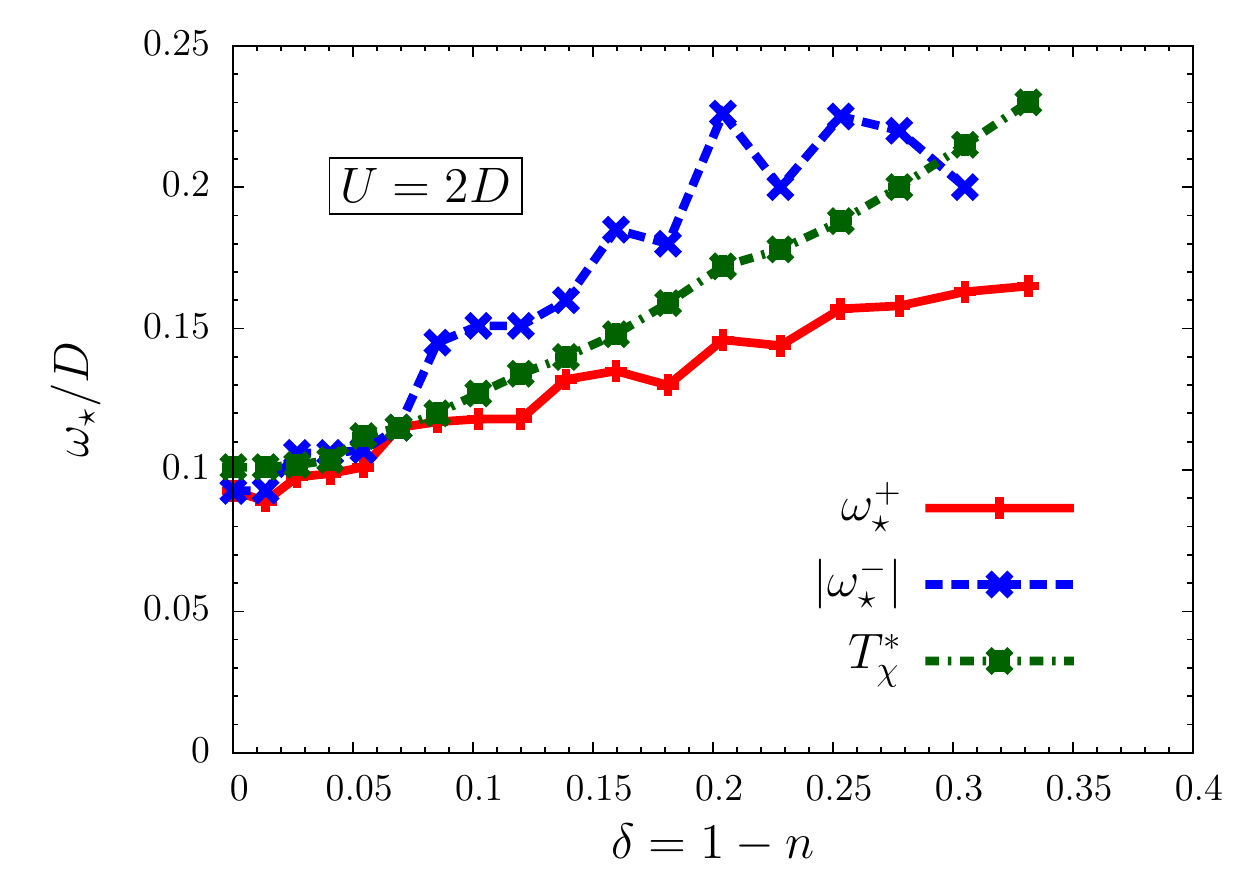}
  \includegraphics[width=\columnwidth]{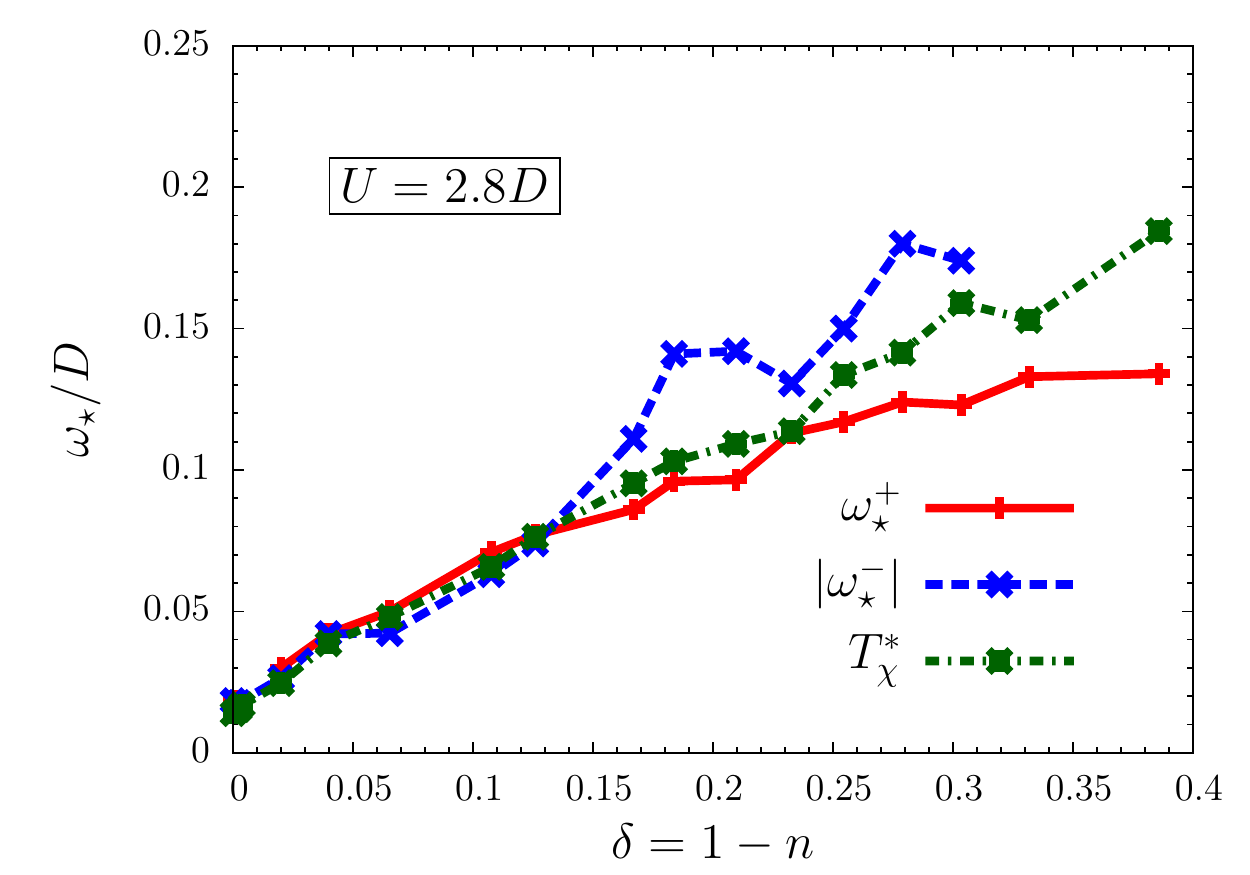}   
  \caption{(Color online) 
    Dependence of the kink positions $\omega_\star^\pm$ and of the scale of
    the magnetic modes on hole doping for $U=2D$ (upper panel)
    	and for $U=2.8D$ (lower panel) obtained from DMFT(NRG).}
  \label{fig:kink_doping}
\end{figure}

Finally, we study the doping dependence of the kinks at fixed interaction.
Generic results are depicted in Fig.\ \ref{fig:kink_doping} for hole 
doping.  As observed in Fig.\ \ref{fig:pos_kink_sus}
the kink positions and energy scale of the spin fluctuations coincide 
$\omega_\star^+\approx |\omega_\star^-| \approx T_\chi^*$ 
for small doping $\delta \lessapprox 0.07$.
For larger doping the particle-hole asymmetry implies that $\omega_\star^+$
and  $|\omega_\star^-|$ differ from each other and hence from $T_\chi^*$.
For hole doping, we find $\omega_\star^+ > T_\chi^* > |\omega_\star^-|$\
but the deviations are rather small. 
Up to an offset, all three energy scales 
depend essentially linearly on doping. The energy scales rise upon 
increasing doping. The two panels of Fig.\ \ref{fig:kink_doping} compare
the doping dependence of the kinks for two different values of $U$.
Clearly,  a larger value of $U$ decreases the energy scale of the kinks
as one would expect for a magnetic energy scale.

At this point, a comparison to experiment is in order. The experimentally best 
studied strongly correlated systems displaying kinks are the superconducting 
cuprates. It is presently still debated whether these kinks are
of phononic\cite{lanza01,mishc09} or of magnetic origin.\cite{eschr06,dahm09}
Our calculation based on DMFT and a semi-elliptic DOS is too far away from
the experimental situation to make a quantitative comparison.
But it is interesting to note that the kink positions observed at low
temperatures in underdoped high $T_c$ materials indeed display
a linear behavior with offset very similar to the one in Fig.\ 
\ref{fig:kink_doping}. Even the numbers are in the experimental range 
\cite{kordy04,kordy06} of about $30$ meV at zero doping
to $120$ meV at $\delta=0.15$ if we assume $D=1.4$eV and $U=4$eV$\approx 2.8D$.
The latter number is too high by a factor 2 compared to the cuprates, 
\cite{kordy06} but this is not astounding in view of the  approximation made 
in the present study.

We emphasize that we consider only the low-energy kinks. These
can be expected to be described with an effective low-energy
single-band Hubbard model which is based on the existence of
the Zhang-Rice singlets.\cite{zhangRice88}
Any statements on the formation of Zhang-Rice singlets
and on dispersion features at higher energies such as ``waterfalls''
which require a multiband description, see for instance Ref.\  
\onlinecite{inoso07b}, are beyond the scope of the present work.
Still, the doping dependence observed here 
is in accord with a magnetic explanation of the kinks in the cuprates.

\section{Summary}
\label{sec:summary}

In this paper, we have  focussed on the physical
origin of kinks in the electronic dispersion 
of strongly correlated electronic systems \textit{without} 
coupling to external bosons.
Exemplarily, we have  studied the doped Hubbard model in 
dynamic mean-field theory at zero temperature.

Three different numerical algorithms have been employed for solving 
the self-consistent impurity problem for the dynamic mean-field  theory
and their results  have been compared.
We have found a very good agreement between all  three
impurity solvers. This is quite remarkable considering their very 
different nature and corroborates the validity of our  evaluations.
We distinguish the two 
numerical renormalization group approaches (NRG and DMRG) which can be applied
directly at zero temperature and the analytical
ENCA which is based on the summation of a large subset of diagrams of
an expansion in the hybridization.

The NRG is an efficient numerical approach for arbitrary temperatures 
and very precise at small
frequencies, but washes out  spectral  details at higher frequencies due to
the logarithmic discretization and the concomitant broadening.
The DMRG is the most resource consuming approach because
it requires a separate run for each spectral frequency.
It is used here as a zero temperature method,
though extensions to finite temperatures are in principle possible.
To access exponentially small frequencies a logarithmic discretization
would be necessary.\cite{nishi04a} For equidistant discretization, however, 
the DMRG exhibits a very good resolution at all frequencies. 
The ENCA is fastest in the computation and
provides data with arbitrary resolution at all frequencies. But it suffers
from Fermi liquid pathologies below the coherence temperature $T^*$
which prevent its reliable application at very small temperatures.

With all these three different methods,
we have studied the kinks in the dispersion of the interacting
electron system away from half-filling. We have established that the appearance
of the kink is linked to dominant spin fluctuations at low energies. In 
particular, the position of the kinks in energy is intimately linked
to the  spin-fluctuation scale $T^*_\chi$ where the magnetic susceptibility
exhibits its maximum. 
Additionally, the occurrence of kinks requires a substantial 
energy separation between the charge
fluctuation scale $U/2-|\mu|$ and $T^*_\chi$, i.e.~$U/2-|\mu|\gg T^*_\chi$.
This is clearly the case in the strongly correlated regime at large values of 
$U\gtrapprox 2D=W$  where $W$ is the bandwidth.
Therefore, the low-lying bosonic modes of
the electronic system modify the electronic dispersion on the scale  $T^*_\chi$
at sufficiently strong electron-electron interaction $U$.

Universality, i.e., the possibility to describe the low-energy  dynamics 
in units of a single energy scale, is only observed close to half-filling
where the scale is set by the spin-fluctuation scale $T^*_\chi$.
For large doping and/or weaker interaction, there is no
clear separation between the spin- and charge-fluctuation scale
and universality is lost.

Finally, we have compared the doping dependence of the kinks in the single-band
Hubbard model  with  
kinks measured with  angle-resolved photoemission in planar cuprates.
In spite of the much higher complexity of the cuprates compared to our
model study, the energy scales are  qualitatively reproduced: Magnetically
induced kinks evolve from 30 meV to about 120 meV with increasing 
hole doping for $U=1.4W$.

In conclusion, we extended the view that emergent modes 
of the electronic system
can generate kinks in the electronic dispersion in analogy to materials with
strong electron-phonon coupling
even away from half-filling,
provided (i) that there is a significant energy separation between the 
high and low energy scale, $T^*_\chi/D\ll 1$ andwhich implies
that we are in the strong coupling regime 
$U/W\gtrapprox 1$, and (ii) that the charge energy scale is much 
larger than $T_\chi^*$. Note that in the studied
strongly correlated system the ratio $T^*_\chi/D$ takes the role
of the ratio of the Debye frequency over the Fermi energy in coupled
electron-phonon systems.
We expect that this a generic feature and very similar results will apply
in many other related systems with clearly separated energy scales 
and strong emergent modes.

\acknowledgments
We acknowledge financial support from the Deutsche Forschungsgemeinschaft under
AN  275/6-2
and supercomputer support by the NIC, FZ J\"ulich under project No.\ HHB00. 
One of us (PG) was supported by a PhD grant of the Studienstiftung des 
deutschen Volkes.






\end{document}